# Haze in Pluto's Atmosphere

A. F. Cheng[1], M. E. Summers[2], G. R. Gladstone[3], D. F. Strobel[4], L. A. Young[5], P. Lavvas[6], J. A. Kammer[5], C. M. Lisse[1], A. H. Parker[5], E. F. Young[5], S. A. Stern[5], H. A. Weaver[1], C. B. Olkin[5], K. Ennico[7].

[1]JHU/APL (andrew.cheng@jhuapl.edu), [2]George Mason University, Fairfax, VA, [3]Southwest Research Institute, San Antonio, TX, [4]The Johns Hopkins University, Baltimore, MD, [5]Southwest Research Institute, Boulder, CO. [6]University of Reims, Reims, France. [7]NASA Ames Research Center, Moffett Field, CA.

**Abstract**

Haze in Pluto's atmosphere was detected in images by both the Long Range Reconnaissance Imager (LORRI) and the Multispectral Visible Imaging Camera (MVIC) on New Horizons. LORRI observed haze up to altitudes of at least 200 km above Pluto's surface at solar phase angles from ~20° to ~169°. The haze is structured with about ~20 layers, and the extinction due to haze is greater in the northern hemisphere than at equatorial or southern latitudes. However, more haze layers are discerned at equatorial latitudes. A search for temporal variations found no evidence for motions of haze layers (temporal changes in layer altitudes) on time scales of 2 to 5 hours, but did find evidence of changes in haze scale height above 100 km altitude. An ultraviolet extinction attributable to the atmospheric haze was also detected by the ALICE ultraviolet spectrograph on New Horizons. The haze particles are strongly forward-scattering in the visible, and a microphysical model of haze is presented which reproduces the visible phase function just above the surface with 0.5 μm spherical particles, but also invokes fractal aggregate particles to fit the visible phase function at 45 km altitude and account for UV extinction. A model of haze layer generation by orographic excitation of gravity waves is presented. This model accounts for the observed layer thickness and distribution with altitude. Haze particles settle out of the atmosphere and onto Pluto's surface, at a rate sufficient to alter surface optical properties on seasonal time scales. Pluto's regional scale albedo contrasts may be preserved in the face of the haze deposition by atmospheric collapse.

1. Introduction

NASA's New Horizons spacecraft made the first reconnaissance of the Pluto-Charon system with closest approach on July 14, 2015, obtaining images and spectra of Pluto and its satellites as well as in situ measurements and other data types (Stern et al 2015). Departure images at high solar phase angles, obtained by the Long Range Reconnaissance Imager (LORRI) on New Horizons, unexpectedly revealed that Pluto's atmosphere is hazy [Stern et al. 2015]. LORRI is a panchromatic visible imager with a bandpass of about 350 nm to 850 nm and a pivot wavelength of 607.6 nm [Cheng et al. 2008]. The pivot wavelength is $\lambda_p = \left(\frac{\int P(\lambda)\, \lambda d\lambda}{\int P(\lambda)\, d\lambda/\lambda}\right)^{1/2}$ where $P(\lambda)$ is the photon detection efficiency across the pass band as a function of wavelength $\lambda$. LORRI is a catadioptric, SiC telescope of 20.8 cm aperture with no focusing mechanism, and it has a field-of-view FOV=0.29° and a pixel resolution IFOV = 4.96 µrad. The detector is a 1024×1024 pixel (optically active region), thinned, backside-illuminated silicon charge-coupled device operated in frame transfer mode.

The haze in Pluto's atmosphere is globally extensive (Stern et al. 2015; Gladstone et al. 2016) and is detected in the New Horizons images by both LORRI and also the New Horizons Ralph imaging instrument (Reuter et al. 2008), as well as in the UV solar occultation by the Alice UV spectrograph (Stern et al. 2008). Haze is imaged in the visible to above 200 km altitude, and is detected by Alice to 300 km altitude. There are numerous embedded layers in the haze, some of which have horizontal extents of several hundreds of km. Many of the haze layers have small thicknesses of only a few km. The haze also has a bluish color at visible wavelengths, suggesting a composition of very small particles, and it is strongly forward scattering. The observed scattering properties of the haze are consistent with a tholin-like composition. Buoyancy waves generated by winds flowing over orography may be related to the origin of the narrow haze layers.

Pluto's atmosphere was definitively discovered by observations of a 1988 stellar occultation (Elliot et al. 1989; Hubbard et al. 1988). The Elliot et al. (1989) light curve exhibited an inflection point attributed to extinction by an atmospheric haze layer. However, this feature could equally well be attributed by them to an atmospheric thermal inversion, affecting the profile of refractive index with altitude without requiring haze (Eshleman 1989; Hubbard et al. 1990). Subsequent multi-wavelength visible and infrared observations of a 2002 stellar occultation (Elliot et al. 2003) provided additional evidence for atmospheric extinction, suggesting the presence of haze in Pluto's atmosphere. The inflection point in the 1988 occultation light curves could be explained either by a haze layer with a scale height of~30 km and with vertical optical depth of ~0.145, or by a steep, near-surface temperature gradient (Elliot and Young 1992). In the former case, the haze layer model for the stellar occultation light curve required a Pluto radius <1181 km, whereas in the latter case the thermal gradient model for the light curve implied a Pluto radius of 1206±11 km (Elliot and Young 1992). New Horizons results showed that the 1988 occultation light curve inflection is primarily explained by the thermal gradient model: the steep near-surface temperature gradient ~2-5 K/km is present, the Pluto radius is 1189.9±0.4 km, and the vertical optical depth of Pluto haze is only ~0.013 (Gladstone et al. 2016).

Pre-encounter photochemical models of Pluto's atmosphere did not predict any high altitude haze region, because the atmosphere above ~15 km altitude was too warm to expect condensation of hydrocarbons or nitriles (Summers et al., 1997; Krasnopolsky and Cruikshank 1999; Zhu et al. 2014). A haze layer might have been predicted only within the lowest several km of altitude above the surface, where the atmospheric temperature becomes low enough that these species would condense. These photochemical/vertical transport models of Pluto's atmosphere were adaptations of Triton atmosphere models (Strobel and Summers 1995): Triton has a 14±1 μbar N2 atmosphere (Gurrola, 1995), similar to Pluto's 11±1 μbar $N_2$ atmosphere measured by New Horizons (Gladstone et al. 2016) and to pre-encounter predictions for Pluto (Lellouch et al. 2015), with methane as a minor constituent in both cases. In both atmospheres, production of hydrocarbons such as $C_2H_2$, $C_2H_4$, $C_2H_6$ from a combination of solar direct and interplanetary scattered Lyman-α radiation was predicted whereas nitriles such as HCN were the product of $N_2$ photochemistry driven by solar EUV radiation. Eddy mixing was predicted to transport these condensable constituents downward to the lower atmosphere where they would condense at the low temperatures near the surfaces.

The New Horizons Pluto encounter showed that Pluto's atmosphere has a markedly different thermal structure from Triton's atmosphere [Elliot et al. 2000, Gladstone et al. 2016]. Pluto's atmosphere is extensively hazy to above 200 km altitude, as was first recognized in three LORRI images obtained at a solar phase angle of 167°, with LORRI pointed only 13° away from the Sun. These three images were 150 ms exposures obtained at MET 299206714, 299206715, and 299206716 on July 14, 2015, at a range of 360,800 km from Pluto. These were the discovery images because they were included in the quick-look dataset and initially returned to Earth on July 18, 2015 with lossy compression. Even with the compression, these images already showed haze layers. The discovery images have been returned with lossless compression and are shown, with the three images stacked, in Fig. 1. The atmospheric haze as seen in Fig. 1 peaks in brightness at the surface with a maximum haze brightness $I/F$ ~ 0.22, where $I$ is the scattered radiance and $\pi F$ is the solar irradiance. The haze brightness decreases with increasing altitude and appears brighter to the north.

The discovery images were acquired within the observation sequence named P_MULTI_DEP_LONG_2 (see Table 1). Table 1 is a catalog of the LORRI observation sequences containing haze observations discussed in this paper, but it is not an exhaustive catalog of LORRI images in which haze was detected. Atmospheric haze is detected not only in departure images at high phase, but also in approach images at low and moderate phase angles, that were acquired prior to the discovery images. Additional LORRI observations showing haze continued after the last observations listed in the table, at greater distance from Pluto and lower image resolution. Finally, Table 1 does not include Ralph observation sequences with haze detections or the Alice solar occultation observations, in which extinction due to haze is also detected.

Table 1 **LORRI haze image catalog (observations discussed in paper)**

| Sequence Name | Start time (UTC) | Range (km) to Pluto | Phase (deg) | Time Relative to Closest Approach (hr) |
|---|---|---|---|---|
| P_LORRI | 7/14/2015 8:14 | 177792 | 19.5 | -3.6 |
| P_MVIC_LORRI_CA | 7/14/2015 11:34 | 17297 | 67.3 | -0.24 |
| P_HIPHASE_HIRES | 7/14/2015 12:04 | 18758 | 148.3 | 0.24 |
| P_MULTI_DEP_LONG_1 | 7/14/2015 15:42 | 193342 | 169.0 | 3.78 |
| P_LORRI_DEP_0 | 7/14/2015 18:26 | 328830 | 167.3 | 6.60 |
| P_MULTI_DEP_LONG_2 | 7/14/2015 19:05 | 360779 | 167.1 | 7.25 |
| P_LORRI_ALICE_DEP_1 | 7/14/2015 21:03 | 458186 | 166.6 | 9.21 |
| P_LORRI_FULLFRAME_DEP | 7/15/2015 03:26 | 775278 | 165.9 | 15.61 |

Range and solar phase angle are for the first image of each sequence

Haze is not detected at Charon, nor was any evidence for an atmosphere found by New Horizons (Stern et al. 2016). A 3σ upper limit for an $N_2$ atmosphere on Charon of 4.2 picobars was obtained from the UV solar occultation observed by ALICE. A 3σ upper limit of $I/F = 2.6 \times 10^{-5}$ for the brightness of any atmospheric haze layer on Charon was obtained from LORRI images, acquired by the P_MULTI_DEP_LONG_2 sequence (see Table 2). The 3σ upper limit to Charon haze brightness is a factor ~7000 smaller than the haze brightness shown in Fig. 1 for Pluto at approximately the same solar phase angle of 167°.

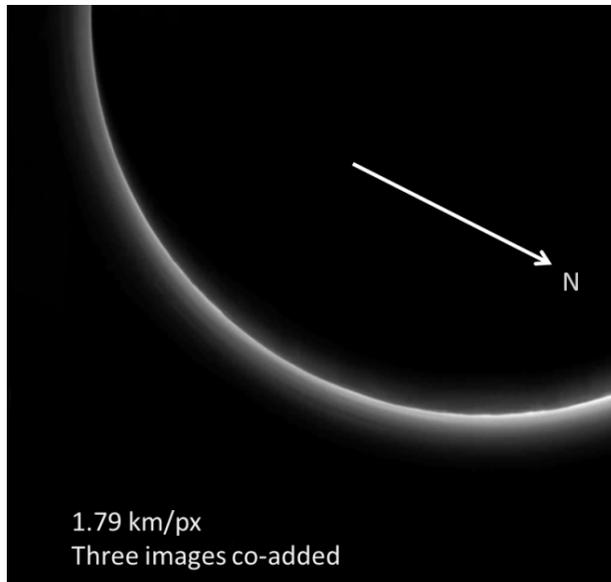

**Figure 1. Pluto haze discovery images. The brightness I/F peaks at the surface at a value I/F~0.22 (near right side of image). Haze layers are visible. N indicates Pluto north. The Sun is to the bottom of the image. Solar phase angle is 167°.**

This paper presents observations of Pluto's haze utilizing mainly LORRI observation sequences (Table 1) with additional information from Ralph images and from the Alice UV solar occultation whose initial results were presented by Gladstone et al. (2016). Section 2 discusses the global distribution of haze with altitude and morphology of haze layers. Section 3 discusses spatial and temporal variation of haze which is brighter in the northern hemisphere. Section 4 presents the strongly forward scattering phase function inferred for Pluto haze. Sections 5 and 6 present microphysical models of the haze, where Section 5 models I/F profiles versus phase angle for spherical particles and for fractal aggregate particles, and Section 6 considers haze production mechanisms and time scales for growth and sedimentation of haze particles. Section 7 presents a model for atmospheric gravity waves from orographic forcing, suggested as a mechanism for producing Pluto's haze layers by Gladstone et al. (2016). Section 8 discusses implications of haze in Pluto's atmosphere.

2. **Distribution of Haze**

Full disk images of Pluto at high phase angle show that the haze extends completely around the limb. Fig. 2 is a stack of four such LORRI images (MET 299236719 to 299236809, each exposure 150 ms) from the sequence P_LORRI_FULLFRAME_DEP at a range of 775,300 km from Pluto and at a solar phase angle 166°.

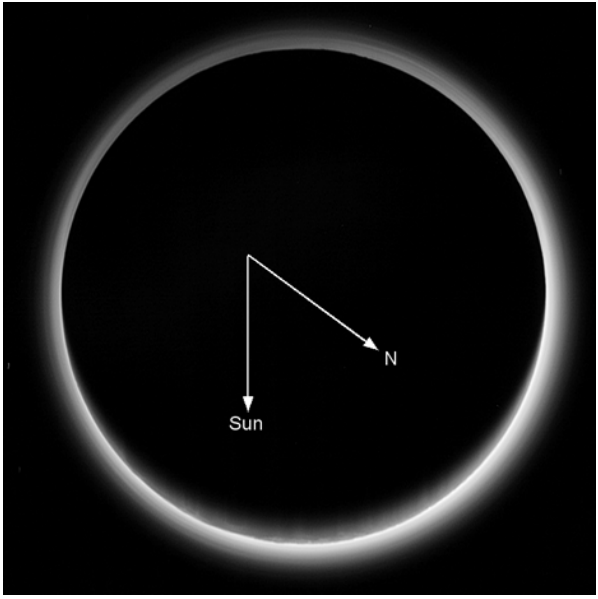

Figure 2. Stack of four images from P_LORRI_FULLFRAME_DEP at 3.85 km/px resolution. Haze is present all around the limb, brightest toward Pluto N and not toward Sun. Layers are seen near 1 o'clock and 7 o'clock positions. Image has been stretched and sharpened. Surface features are seen within the limb of Pluto, and topography at the limb.

At an image resolution of 3.85 km/pixel, haze layers are seen all around the limb, although layers are seen more distinctly at some position angles around the limb than at others. Notable in Fig. 2 is that the haze is brightest toward Pluto north (marked by N in Fig 2) and not in the direction towards the Sun. A distinct left-right brightness asymmetry is evident in the image. The sub-observer longitude and latitude, or Pluto (longitude, latitude) at the center of the observed disk, is (288.7°, -43.9°). The sub-solar (longitude, latitude) is (91.1°, 51.6°), putting the Sun almost directly behind Pluto, but displaced downward in the image.

Fig 3 shows the height distribution of the haze derived from the image stack of Fig 2. The azimuthally averaged I/F versus radius is obtained as in aperture photometry, using concentric circular apertures of successively larger radii from Pluto center. The I/F versus radius is found from the average DN per pixel within each of the annuli between successive apertures. This curve has been differentiated to find the scale height. The center of Pluto is found by fitting a circle to the limb, with limb radius placed at the maximum brightness gradient. The adopted Pluto radius (from fitting to high phase departure images of P_MULTI_DEP_LONG_1) is 1190±1 km, consistent with the radio occultation radius 1189.9±0.4 km (Gladstone et al. 2016) and the radius 1187±4 km from approach images (Stern et al. 2015).

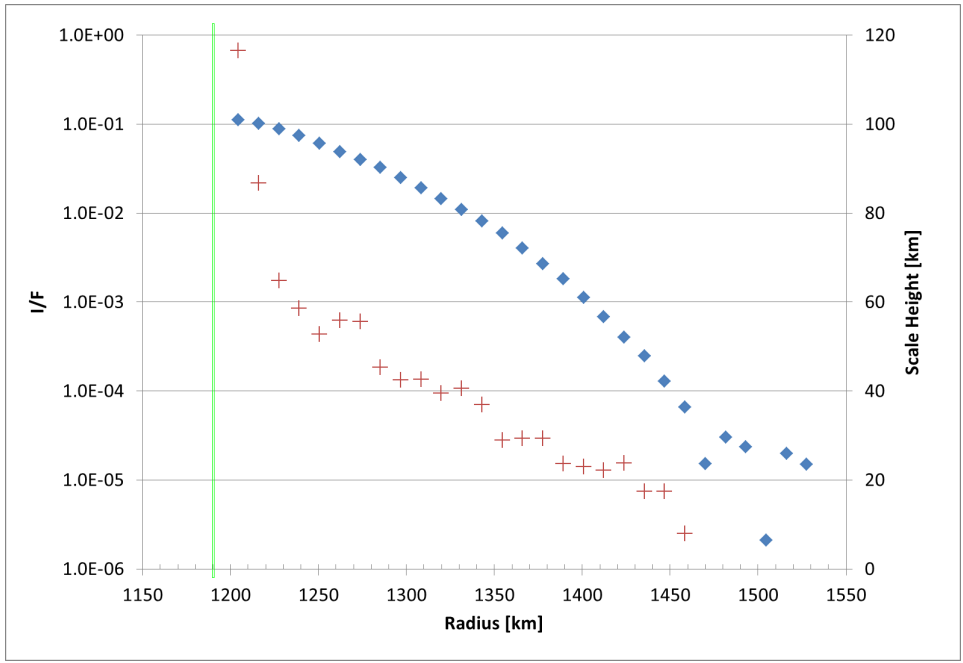

Figure 3. Azimuthally averaged I/F from image sequence P_LORRI_FULLFRAME_DEP, blue diamonds (left axis). Scale heights, red plus symbols (right axis). Green line, Pluto's surface radius 1190 km

Fig. 3 shows an approximately exponential profile for I/F. Plotted on a log scale, the I/F profile is concave downward indicating a decrease in scale height with increasing altitude, up to 200 km height. The noise floor is reached in this observation at ~1450 km Plutocentric distance, because of stray light (the Sun is only 14° off the camera boresight). Below 1400 km, the uncertainty in I/F is within the size of the symbol plotted.

Haze layers can be discerned both in Fig. 1 and in Fig. 2. They are approximately horizontal, and they can be traced over lateral extents of hundreds of km. The layer structures are brought out in high resolution images obtained by LORRI and MVIC near closest approach. Fig. 4 is just such a high resolution (340 m/px), panchromatic MVIC image [MET0299181303, after stretch and crop] obtained at high phase angle of 147°. At this phase angle, haze layers are easily visible and have relatively similar brightness to Pluto's surface.

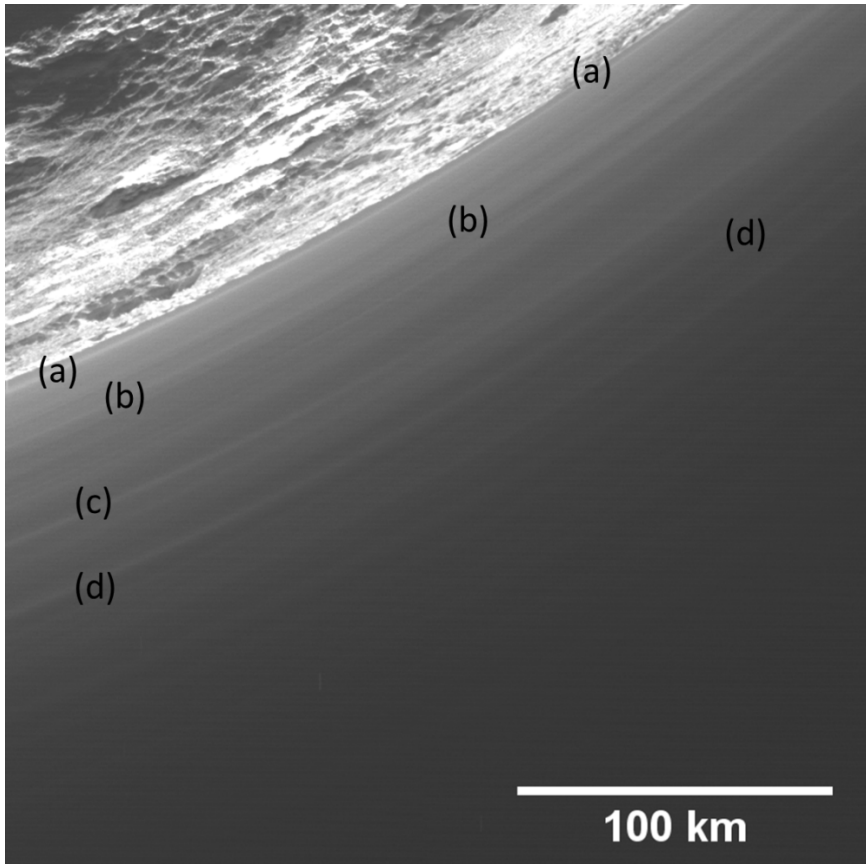

Figure 4. Haze layers change thickness (a), merge/split (b), or appear/disappear (c). A dark lane at altitude 72 km is marked by (d), near minimum of atmospheric buoyancy frequency. MVIC MET0299181303 at phase 147°, with scale bar.

Fig. 4 shows that haze layers merge, separate (divide into thinner layers), change in thickness, or appear and disappear, when traced around the limb. The minimum haze layer thickness is about 1 km, but some layers appear over 10 km thick. The layers are predominantly horizontal structures. Also seen in Fig 4 is a dark lane of low haze brightness, at altitude 72 km. Near this altitude, the atmospheric buoyancy frequency reaches its minimum value, and in the gravity wave model of Section 7, the vertical wavelength of gravity waves reaches its maximum value.

Fig 5 is the highest resolution image of haze layers, obtained by LORRI as a ride-along image during the spacecraft scan to obtain the MVIC image of Fig. 4 (MVIC is a time-delay integration imager). This slew necessitated a very short 10 ms exposure time for LORRI to minimize image smear, causing the image to be under-exposed with row-striping artifacts. Fig 5 shows the same region of Pluto as seen in Fig 4 with MVIC. In Fig 5 the lowest altitude haze layer, which is indicated by (a) in Fig 4, is not resolved into still finer sub-layers but is a single, bright layer with width varying from 1 to 3 km. This lowest layer is found at 3 km altitude at the location of Fig 5, and it is the same bright layer as shown by Gladstone et al. (2016), descending from ~5 km above the surface, down to the surface, over a horizontal distance of 600 km. The LORRI image of Fig 5 was acquired in the sequence P_HIPHASE_HIRES (see Table 1) with resolution 93 m/px at phase 148°.

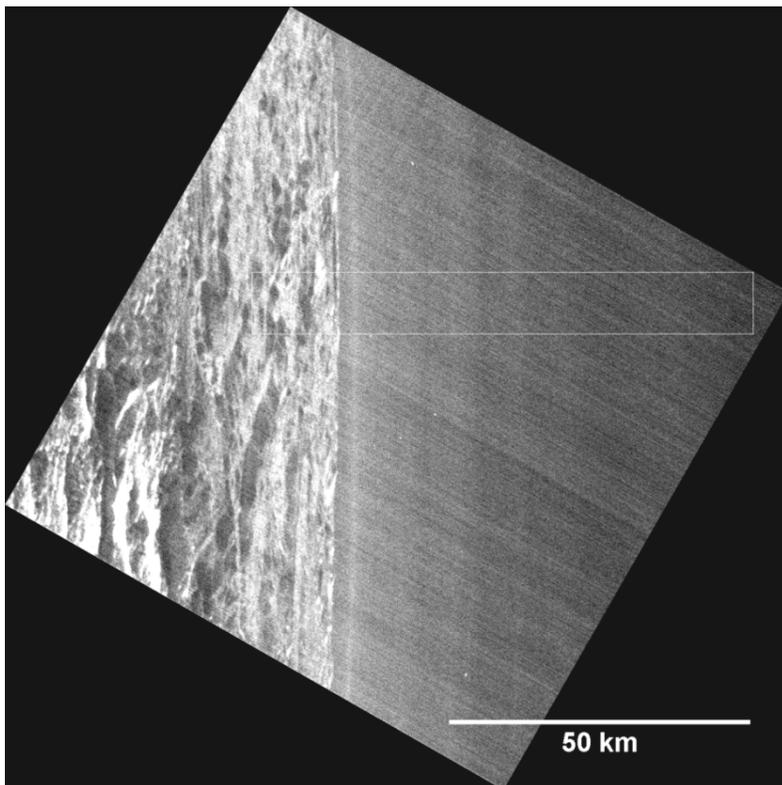

**Figure 5. Image MET0299181359 from LORRI P_HIPHASE_HIRES. The highest resolution haze image, showing minimum layer thickness ~1 km. Phase angle 148°. See text for explanation of box. A 50 km scale bar indicated.**

About 20 haze layers are detected in Pluto's atmosphere (Fig 6). This same region of the limb is seen with prominent and distinct layering also in Figs.1 and 2, near the 7 o'clock position. The higher resolution image of Fig 6, at 0.96 km/px, also shows limb topography at amplitudes of several km, as well as crepuscular rays (Stern et al. 2015) in the form of vertical dark shadows cast by topography, within the limb of Pluto. The location at the limb of Pluto indicated by the yellow star in Fig 6 corresponds to the location marked by the yellow star on the inset basemap of Pluto. This is an equatorial location, near the "tail" of Cthulhu, the prominent equatorial dark region on Pluto west of the bright, heart-shaped region Tombaugh Regio (all Pluto names are informal). Sputnik Planitia, the left side of the heart shape, is a nitrogen ice cap which sits in a depression 3 to 4 km below surrounding rugged uplands, located both to the east and to the west.

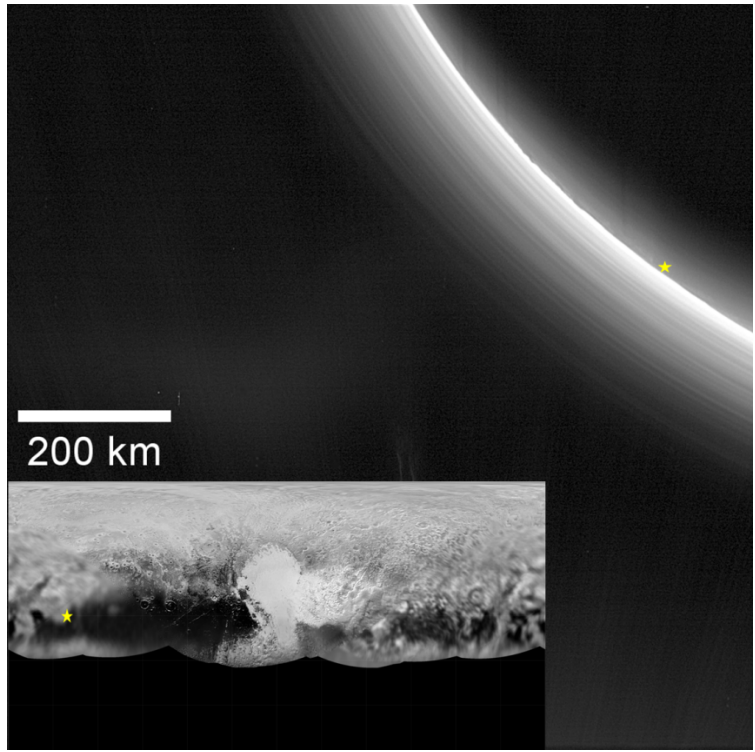

**Figure 6. stack of two images MET0299194661 and 299194671 from MULTI_DEP_1. About 20 haze layers detected at 0.96 km/px. Inset shows Pluto basemap, with stars indicating corresponding Pluto locations. Sun is to bottom.**

### 3. Spatial and Temporal Variations of Haze

Fig 7 shows an "unwrapped" and edge-filtered haze image, calculated from a mosaic of 12 images obtained by sequence P_MULTI_DEP_LONG_1 at 0.96 km/px and at 169° phase. In the unwrapped image, the haze brightness, as a function of radius and clock angle around the limb of Pluto, is re-plotted (after edge filtering) with radius on the vertical axis and clock angle on the horizontal axis. The radius range shown extends to 150 km altitude above the surface. The edge filter is a Roberts cross filter as implemented in IDL, which estimates the spatial gradient of an image from the sum of the squares of the differences between diagonally adjacent pixels. This filtering emphasizes the haze layers to show their global distribution. In Fig 7, the image mosaic seams are also emphasized by edge filtering and appear as thin, white tilted lines.

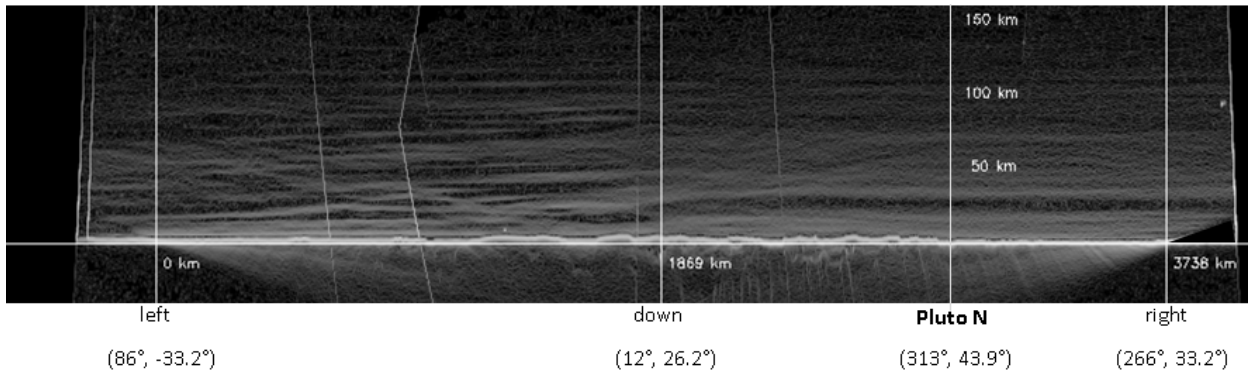

**Figure 7** Unwrapped, edge-filtered mosaic of 12 images MET 299194429 to 299194729 showing haze layers; with altitude above the surface as vertical axis and clock angle around the limb as horizontal axis. Thin, tilted white lines are mosaic seams. A distance scale is shown along the surface at zero altitude. Pluto (longitude, latitude) indicated for limb locations at LORRI image left, image down, and image right, as well as northernmost limb (Pluto N). More disturbed layers are seen to left side, see Figure 8.

As in Fig. 4, about 20 haze layers can be identified in the unwrapped image of Fig. 7. Layers are mainly horizontal, but a few are sloping toward the surface. Individual layers can be traced to over 1000 km horizontal extent, but not around the entire imaged portion of the limb (4289 km or 206.5° clock angle extent). More disturbed layers are shown on the left side of Fig 7, where layers are more often seen to appear, disappear, or slope to the surface and merge with one another. The more disturbed layers occur over geographic locations that are identified in Fig 8, which gives limb locations on a Pluto base map.

Also in Fig 7 a generally dark band can be traced across the entire unwrapped image around 75 km altitude, indicating a suppression of layering at that altitude. The minimum of Brunt-Vaisala frequency in the Pluto atmosphere is found near that altitude (see Fig. 24). In the gravity wave model of Section 7, the vertical wavelength of gravity waves also reaches its maximum value near that altitude, leading to a predicted suppression of layering. Another, even wider dark band is evident near 30 km height, but only in the right side of Fig. 7, which corresponds to the more northern latitudes and longitudes from roughly 360° to 250° near the right side edge of the unwrapped image.

Fig. 8 shows the (longitude, latitude) positions on Pluto seen at the limb for four of the observation sequences in Table 1, as calculated from the reconstructed spacecraft trajectory. Finite distance of the spacecraft from Pluto is accounted for, so the observed limb is not a great circle, but a slightly smaller one.

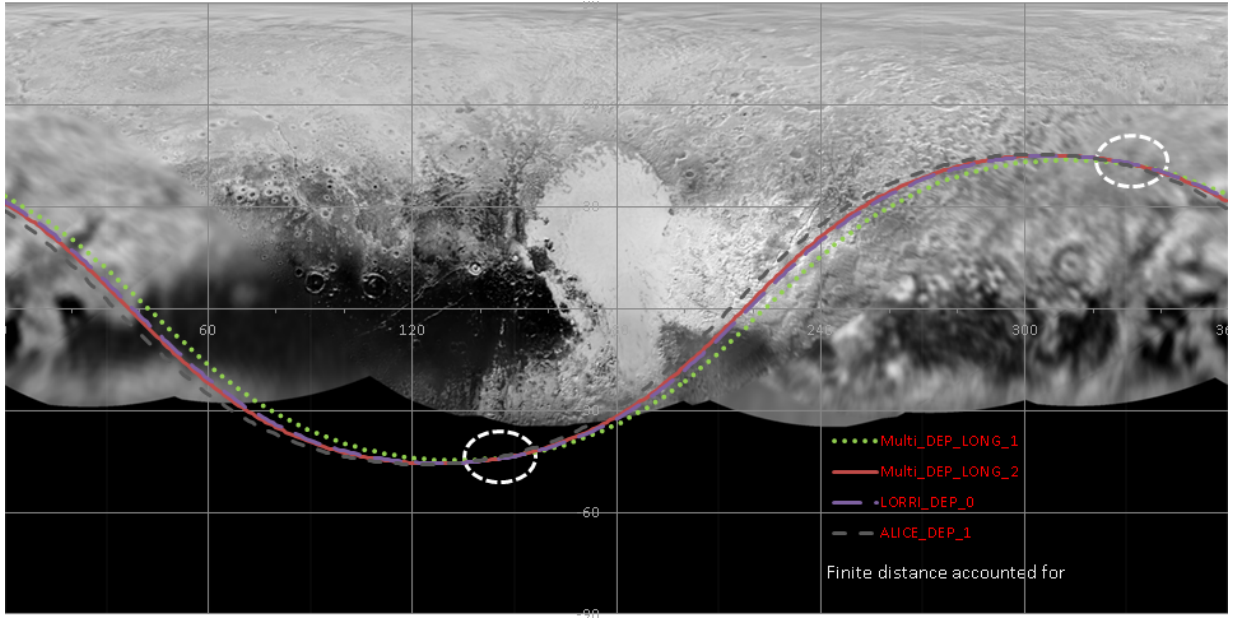

**Figure 8.** Locations seen at the limb, plotted on the Pluto base map for the four departure observation sequences of Table 1. Where limb traces intersect (white ovals), haze over the same locations on Pluto was observed at different times.

Fig. 8 shows that the limb trace for the P_MULTI_DEP_LONG_1 sequence, which obtained the images used in Fig. 7, passes just to the south and east of Sputnik Planum. However, that particular portion of the limb was not imaged. The P_MULTI_DEP_LONG_1 sequence was designed as a 3×4 mosaic, which was not large enough to cover the entire disc of Pluto, because additional time and downlink resources could not be committed. The portion of the limb covered in the unwrapped image (Fig. 7) is roughly the longitude range from 0° to 90° on the base map and then from 250° to 360°. The more disturbed layers in Fig 7 occur in the 0° to 90° longitude range, at generally low latitudes where the limb trace crosses the tail of Cthulhu, the brighter uplands to the north and south, and a portion of Pluto not imaged by New Horizons.

In addition to the study of spatial distributions of haze and haze layers, it is important to investigate temporal variations of atmospheric haze to constrain possible formation mechanisms. The New Horizons departure images, at sufficiently high resolution to characterize layering, covered a time base of several hours (Table 2). However, the New Horizons observations of haze above the limb of Pluto, obtained at different times by different observation sequences, also measured haze over different locations as shown by the limb traces in Fig. 8. To study temporal variations, it is necessary to measure haze over the same locations on Pluto at different times. The intersection points of different limb traces in Fig. 8 then indicate the Pluto locations which were observed at the limb in more than one observation sequence.

A temporal variation search was performed by comparing I/F profiles extracted over the intersection points indicated in Fig. 8, that is, observed at different times over the same geographic locations on Pluto. Table 2 shows four such intersection points, three in the northern hemisphere and one in the southern hemisphere. The columns in Table 2 give the longitude and latitude of intersection points (some of which were not imaged because the sequenced image mosaics did not cover the entire limb), clock angles of these locations around the limb (positive CW, measured relative to Pluto North), and time intervals between observations of the intersection point. The cases where comparisons are made to search for temporal variation are shown as bold in Table 2. Three comparisons are made in the north, at 1.97, 3.46 and 5.43 hour time intervals, and one comparison is made in the south, at a 2.61 hour time interval. The reference angle position to Pluto North in the LORRI images, from which clock angles were measured, is given in Table 3 (measured CW from image UP). For all the image sequences in Table 2, the Sun is toward image DOWN.

**Table 2 Temporal Variation Search using Limb Trace Intersections**

| Intersections | Lon | Lat | Clock Angle in P_MULTI_DEP_LONG_1 | Clock Angle in P_MULTI_DEP_LONG_2 | Clock Angle in ALICE_DEP_1 | Clock Angle in P_LORRI_DEP_0 | Time interval |
|---|---|---|---|---|---|---|---|
| P_MULTI_DEP_LONG_1 & P_MULTI_DEP_LONG_2 | 149 | -43.5 | 191.6* | 195.7* | | | |
| | 334 | 42.0 | **15.5** | **19.7** | | | 3.46 hr |
| P_MULTI_DEP_LONG_1 & ALICE_DEP_1 | 142 | -44.3 | 186.5* | | 193.5 | | |
| | 326 | 43.2 | **9.5** | | **16.6** | | 5.43 hr |
| P_MULTI_DEP_LONG_2 & ALICE_DEP_1 | 132 | -45.5 | | 183.4* | 186.3 | | |
| | 314 | 45 | | **4.9** | **7.8** | | 1.97 hr |
| ALICE_DEP_1 & P_LORRI_DEP_0 | 133 | -45.4 | | | **187** | **183.2** | 2.61 hr |
| | 315.5 | 44.9 | | | 9 | 5 | |

All angles in degrees. *located in portion of limb that was not imaged.

**Table 3 LORRI image geometry**

| Sequence Name | Resolution (km/px) | Phase Angle | Pluto N angle |
|---|---|---|---|
| P_MULTI_DEP_LONG_1 | 0.960 | 169.0° | 126° |
| P_LORRI_DEP_0 | 1.632 | 167.3° | 127° |
| P_MULTI_DEP_LONG_2 | 1.791 | 167.1° | 127° |
| ALICE_DEP_1 | 2.274 | 166.6° | 120° |

Pluto N angle measured CW from image UP

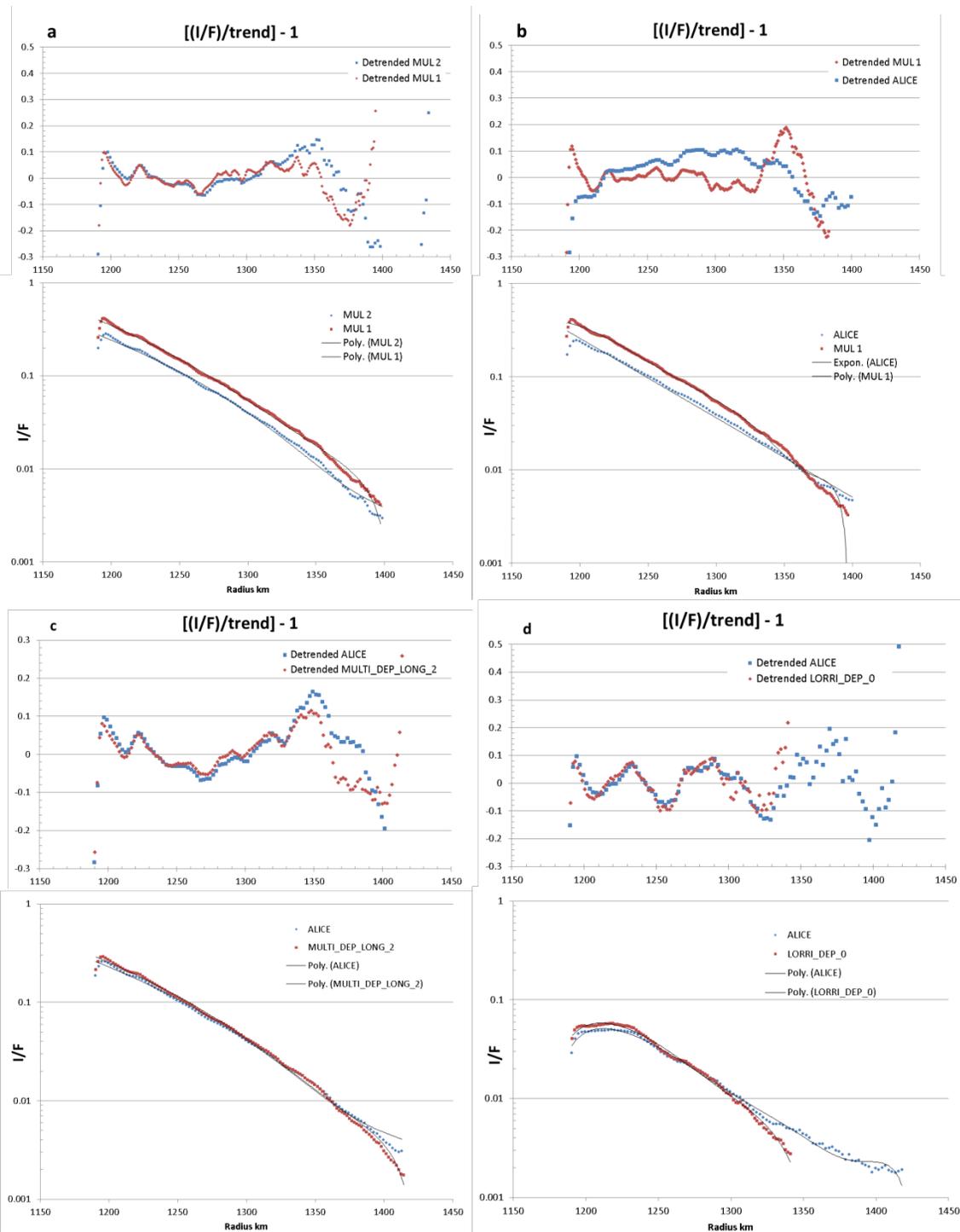

**Figure 9.** Temporal variation search, showing changes in I/F profiles, but not in layer positions. Panel a) comparison of P_MULTI_DEP_LONG_1 with P_MULTI_DEP_LONG_2; b) comparison of the former with ALICE_DEP_1; c) comparison of ALICE_DEP_1 with P_MULTI_DEP_LONG_2; d) comparison of ALICE_DEP_1 with LORR_DEP_0, profile truncated at edge of image frame. Fitted trend lines (exponential or polynomial) shown as black lines. Layer positions agree very well for earlier and later observations in all four panels, even with changes in the overall shape of I/F profiles.

Fig. 9 shows the four comparisons of I/F profiles obtained over the same Pluto locations at different times, in the four panels **a** through **d** of Fig 9. For each comparison, the two detrended I/F profiles are plotted above the two I/F profiles. Panel **a** shows the comparison of P_MULTI_DEP_LONG_1 with P_MULTI_DEP_LONG_2 in the north at longitude 334° and latitude 42°, over a 3.46 hour time interval. The two profiles are in the lower portion of panel **a**, with the trend lines plotted in black; both are cubic polynomial fits. The detrended profiles are in the upper portion of panel **a**, where fractional deviation from the trend is plotted $\left[\left(I/F\right)/trend\right] - 1$.

Panel **b** shows the comparison of P_MULTI_DEP_LONG_1 with ALICE_DEP_1, where in this case the ALICE profile is detrended with an exponential. Panel **c** compares P_MULTI_DEP_LONG_2 with ALICE_DEP_1, where both trend lines were cubic polynomials. Panel **d** compares ALICE_DEP_1 with P_LORRI_DEP_0, where the trend lines were $6^{th}$ order and $5^{th}$ order polynomials, respectively.

The I/F comparisons in Fig 9 (lower portions of each panel) indicate changes in haze brightness between the earlier and later observations. In panel **a**, between P_MULTI_DEP_LONG_1 and P_MULTI_DEP_LONG_2 there was an overall decrease of 34% in haze brightness. A similarly large brightness decrease is shown in panel **b** between P_MULTI_DEP_LONG_1 and ALICE_DEP_1 at low altitudes up to ~150 km, although there is also a change in the haze scale height above 150 km altitude, such that the two observed profiles cross each other above 1350 km radius. A similar change in haze scale height or in I/F profile shape is seen in panel **d**, where a scale height change occurs above 1300 km radius.

However, the haze brightness differences at low altitudes, seen in all four panels of Fig 9, are most likely not due to temporal variations. In all four cases, the brightness change at low altitude is consistent with the small differences in solar phase angle of the observations (Table 3) assuming that the haze is strongly forward scattering. That is, the haze was brighter at low altitude if observed at higher phase angle. It is implausible that the >30% overall brightness changes over more than 100 km altitude, as seen in panels **a** and **b**, are temporal variations. However, the haze scale height or I/F profile shape changes are not explained by a strongly forward scattering phase function, and these changes are attributed to temporal variations.

Also compared in Fig 9 are detrended I/F profiles, in which the presence of haze layers can be seen as wave-like features. The haze layers typically exhibit fractional deviations in I/F of a few to several percent. A striking result shown in all four panels of Fig 9 is that the haze layer altitudes correspond very well between earlier and later observations, over time spans from 1.97hr to 5.43 hr. It is concluded that no evidence is found for consistent motions of layers or changes in layer heights over time spans of several hours. The apparently diverging amplitudes of the oscillations in the detrended profiles at high altitude above 1350 km radius (e.g., in panel **a**) do not indicate growth in haze layer amplitude, but reflect poorly fitting trend lines. The gravity wave model of haze layering (section 7) predicts that wave amplitudes reach saturation already by several km altitude.

Finally it is noted that three of the comparisons in Fig 9, those in panels **a** through **c**, were made at northern latitudes over sunlit limb, whereas one comparison (panel **d**) was made at longitude 133° and latitude -45° in the southern hemisphere over the night side limb. The conclusions are the same for all four cases, that temporal variation is seen as haze scale height changes above 100 km altitude (or I/F profile shape changes) but that no evidence is found for vertical motions of haze layers.

Fig. 10 shows three I/F profiles from similar equatorial locations at different times given in Table 2: blue is at (longitude, latitude) = (43°, -0.5°); red at (32°, 5°); green at (33°, -0.2°). There are consistent decreases in brightness at low altitudes over the three successive observations, but the brightness at 150 km altitude is stable. Again this consistent, low altitude brightness decrease is most likely explained by a forward scattering phase function with a changing phase angle of observation, and not by a temporal variation. The observations occurred at progressively smaller phase angles (Table 3), such that haze is brighter at low altitude when observed at higher phase angle, indicating a strongly forward scattering phase function (next section).

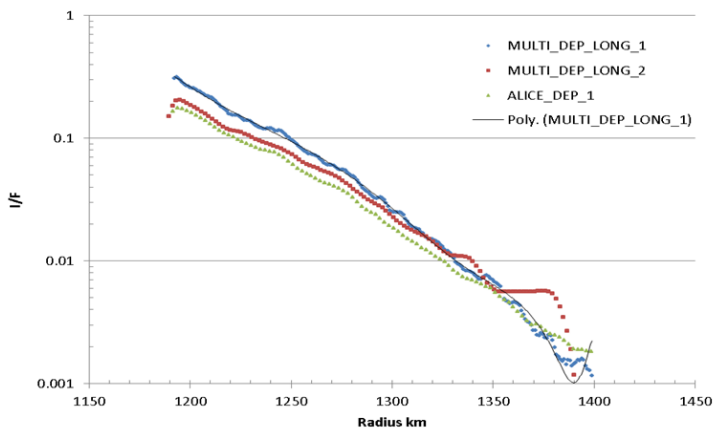

Figure 10. Comparison of equatorial I/F profiles over several hour time interval.

Comparisons of layer features I Fig. 10 are complicated by the substantial changes in resolution over the time span and by the different Pluto locations. However, there are some common features, like a broad brightness minimum in the radius range from ~1210 to ~1220 km and layer maxima around ~1240 km and ~1280 km. The bright feature at 1380 km in the P_MULTI_DEP_LONG_2 profile may be due to poor stray light background removal.

To further quantify the spatial distribution of haze and haze layers, Fig 11 compares profiles from the same observation sequence P_MULTI_DEP_LONG_1 obtained over the most northern latitudes imaged (latitude 44°) and over equatorial latitudes at the same resolution of 0.96 km/px and importantly at constant solar phase angle. The left panel of Fig 11 shows that **haze I/F is systematically greater over the northern mid-latitudes than over the equator**, by factors of 2 to 3. The corresponding comparison of detrended profiles from northern region (lon, lat) = (312°, 44°), versus the equatorial region (43°, -0.5°), from the same time and at the same image resolution, shows that **layers are more distinct in the equatorial region than in the northern region,** as is also evident from examination of the profiles. The equatorial haze layers from the P_MULTI_DEP_LONG_1 sequence are also shown in Fig 6. Both trend lines in Fig 11 are 6$^{th}$ order polynomials. It is difficult to identify correspondences in layer features between northern and equatorial profiles, with some exceptions like the layer just above the surface and three layers at ~1340 km, ~1350 km, and ~1360 km. The haze scale height is also greater in the north, at radii below ~ 1350 km.

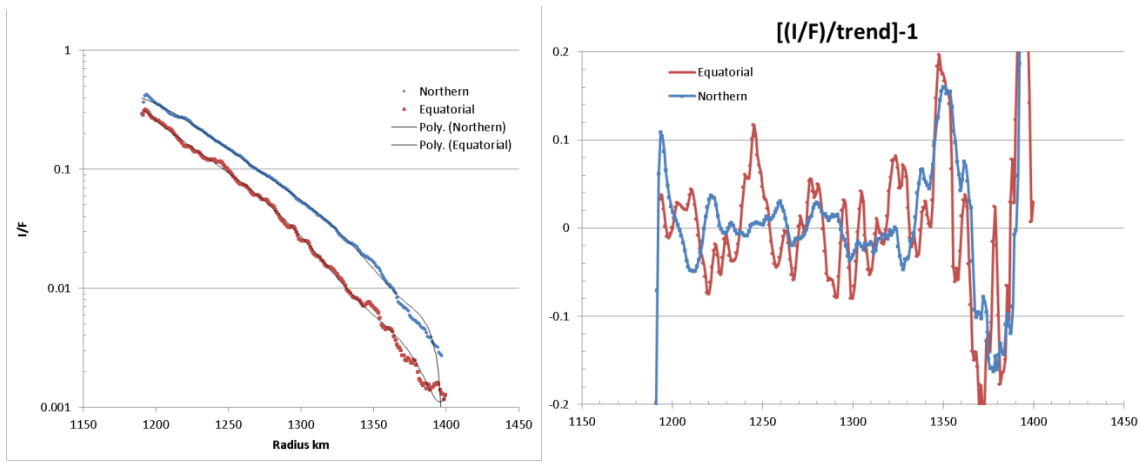

**Figure 11 (left panel) Comparison of I/F profiles from same observation sequence P_MULTI_DEP_LONG_1, equatorial (red) versus northern (blue) near latitude 44°. Northern profile is systematically brighter; equatorial profile has more distinct layering. (right panel) Comparison of detrended I/F, showing generally poor correspondence of layer positions except near 1350 km radius. Differences in haze scale height are also seen between the two profiles at left.**

The spatial distributions of haze and layering from visible imaging are summarized as follows: the haze is brighter in the northern latitudes than over the equator, but layering is more distinct over the equator. This is also seen in Fig. 2, where the haze is brightest around the 4 o'clock position to the north, but layering is most distinct around the 1 o'clock and 7 o'clock equatorial positions. Haze with distinct layers over the equatorial limb is also shown in Figs. 1 and 6. In the unwrapped image of Fig 7, layering is more distinct, and more layers can be discerned, on the left side (equatorial latitudes) than to Pluto north. In general, the maximum of haze brightness is reached several km above the surface.

The result of the temporal variation search is summarized as follows. No motion of layers is detected, but changes in brightness I/F are seen versus height over the same geographic locations, observed over time spans of up to 5.43 hours. The brightness changes at the lowest altitudes correlate to solar phase angle, indicating a strongly forward scattering haze (next section), but changes in the shape of I/F profiles (e.g., Figs 9b and 9d) indicate temporal variations in haze.

## 4. Forward Scattering Phase Function of Haze

The haze in Pluto's atmosphere is optically thin, with the brightness approximated as

$$\frac{I}{F} \cong P(\theta) \frac{n\sigma_s}{4} \sqrt{2\pi RH} \qquad (1)$$

(Gladstone et al. 2016) where $P(\theta)$ is scattering phase function at phase angle $\theta$, $n$ is number density of haze particles, $\sigma_s$ is the scattering cross section, $R$ is Pluto radius, and $H$ is scale height. The phase function is a key diagnostic of physical properties including size, shape, and composition (refractive index). The brightness versus phase angle is considered here, by comparing $I/F$ obtained in back scatter and forward scatter geometries as listed in Table 1.

As haze can be accurately measured only off the limb of Pluto, these observations were made over various locations on Pluto during the 19 hour time span of Table 1. The phase function was compiled using only haze observations over the day side limb at similar northern latitudes (>40°).

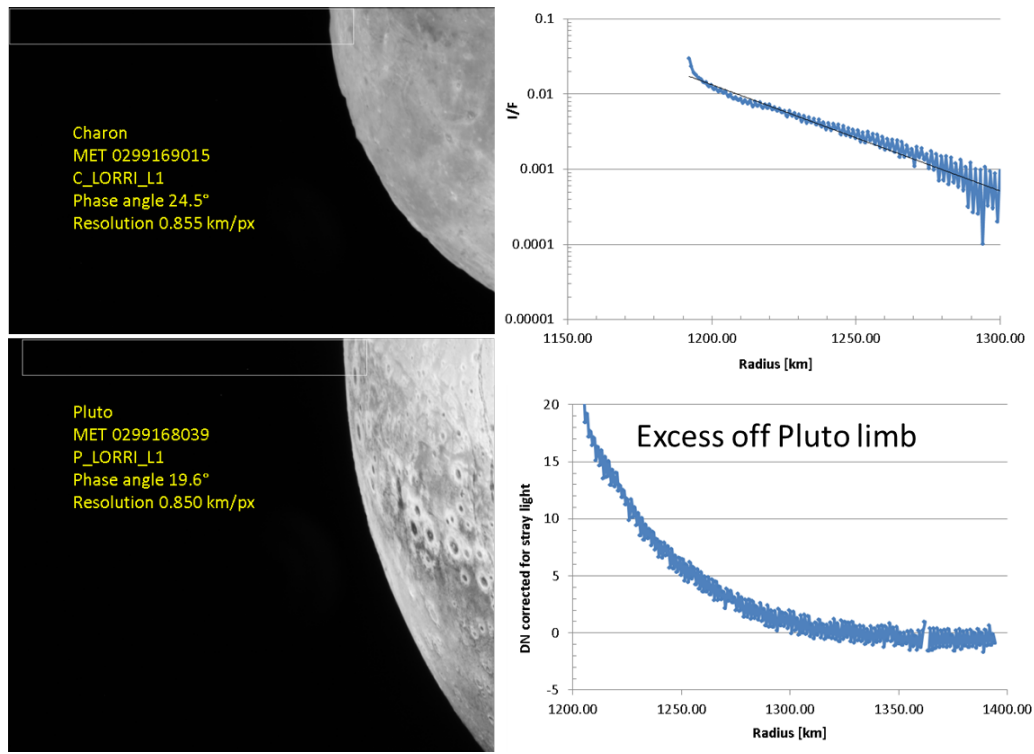

Figure 12. Pluto and Charon imaged near 20° phase off day side limb in similar geometries. Charon profile from stack of 299169015 & 299169016 used to remove stray light from Pluto profile. Boxes indicate where brightness profiles were extracted. These are row profiles, vertically averaged over columns in the box. Exponential fit is shown to I/F profile, indicating a brightness scale height of 31 km

**Haze at 20° phase.** Pluto and Charon were both imaged in similar back scatter viewing geometries and at similar image resolution by the P_LORRI and C_LORRI sequences before closest approach (Table 1). At the low solar phase angles ~20° for these observations, the haze off the limb of Pluto is much fainter than the sunlit surface of Pluto, and the instrumental stray light from the bright limb of Pluto must be removed to measure haze. Fig. 12 shows images of Charon and Pluto from P_LORRI and C_LORRI, obtained in similar viewing geometries, where the Charon image was used to remove stray light from the Pluto image, and the excess brightness in the Pluto image was characterized and attributed to haze. This method takes advantage of the absence of an atmosphere at Charon as well as absence of haze (Stern et al. 2016). A brightness profile is extracted from the indicated box on the Charon image (upper left, Fig. 12) and is scaled by a constant factor to match the brightness (average DN) seen within the sunlit area inside the box on the corresponding Pluto image (lower left, Fig. 12). The scaled Charon profile is subtracted from the Pluto profile to remove stray light.

An excess of DN (lower right, Fig. 12) is seen off the limb of Pluto compared with Charon, yielding an $I/F$ profile (upper right, Fig. 12) with an exponential scale height of 31 km.

The inferred I/F of haze from Fig. 12 shows a sharp drop with altitude just above the surface, whereas the I/F profiles in Figs. 9-11 all show that $I/F$ rises to a peak several km above the surface. The sharp drop in $I/F$ at low altitude is likely attributable to incomplete stray light removal, because of the brightness of Pluto's limb compared to the brightness farther inward, versus the more uniform brightness of the Charon limb.

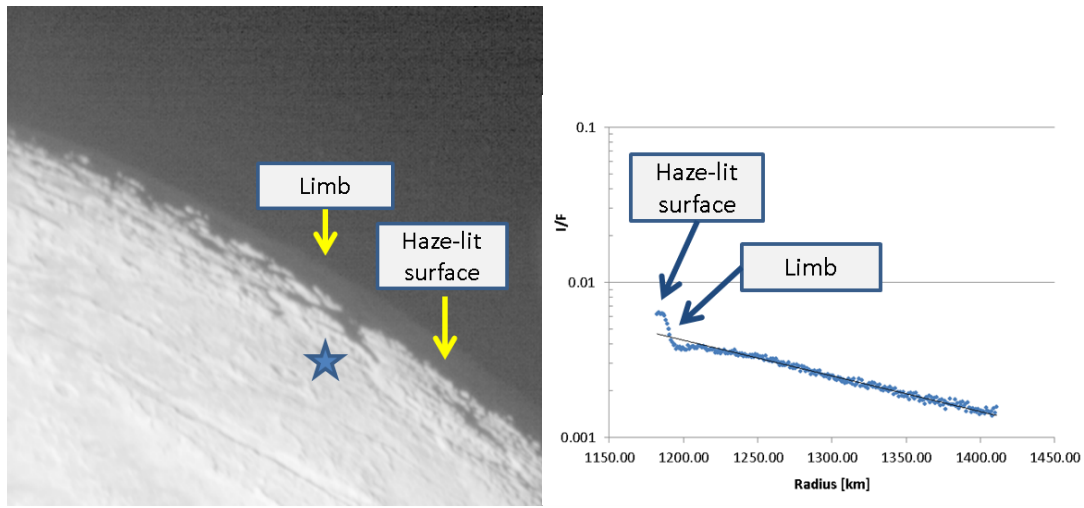

**Figure 13 Night side haze and haze-lit surface, 20° phase. Log stretch of MET0299167703. Scaled Charon profile from stack of 299159015 & 299169016 subtracted from Pluto profile obtained normal to limb over the blue star**

In the same observation sequence P_LORRI, images were also obtained off Pluto's night side limb. These images in Fig. 13 also showed areas of Pluto beyond the terminator and in the night side, that is, areas not directly illuminated by the Sun, but illuminated instead by scattered sunlight from atmospheric haze. Such haze-lit surface is indicated in Fig. 13. Haze above the night side limb is fainter than over the day side by a factor ~4,.

**Haze at 67° phase** The approach sequences P_MVIC_LORRI_CA (Table 1) and C_MVIC_LORRI_CA acquired both Pluto and Charon images only minutes apart in nearly the same viewing geometry, at 67° phase angle. A brightness profile from the Charon image was used to remove stray light from the Pluto profile, assuming absence of haze at Charon (Stern et al. 2016). Fig 14 shows the Pluto image MET0299179658 at resolution 86 m/px, with a box where a column averaged, row profile of brightness is extracted. The box location was chosen to avoid stray light artifacts (haze and artifacts are not visible in the stretch used for Fig. 14). The similar Charon image MET0299180406 (not shown) at resolution 157 m/px was used to remove stray light with a similar box of the same size. As before, the Charon profile was scaled by a constant factor to match the average DN values of the sunlit limb areas within the boxes on the Charon and Pluto images.

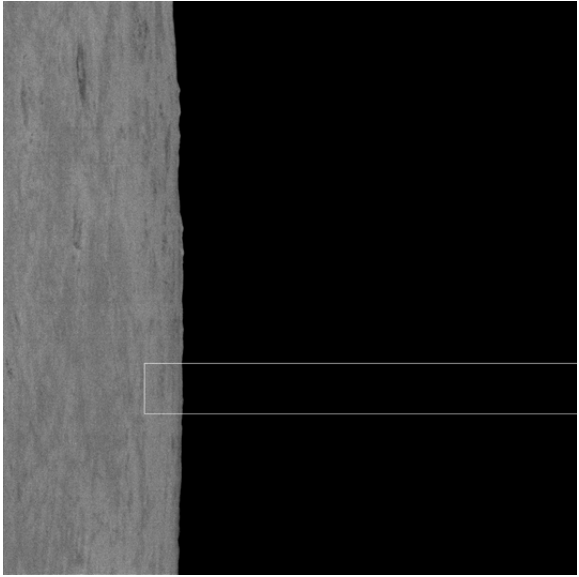

Figure 14 Pluto image MET0299179658 from P_MVIC_LORRI_CA used for haze I/F profile at 67° phase angle.

Fig. 15 (left panel) shows the Pluto DN profile off the limb and the corresponding scaled Charon profile. An excess of DN is seen at Pluto, yielding the $I/F$ profile at 67° phase angle (right panel, Fig. 15) with 39 km exponential scale height.

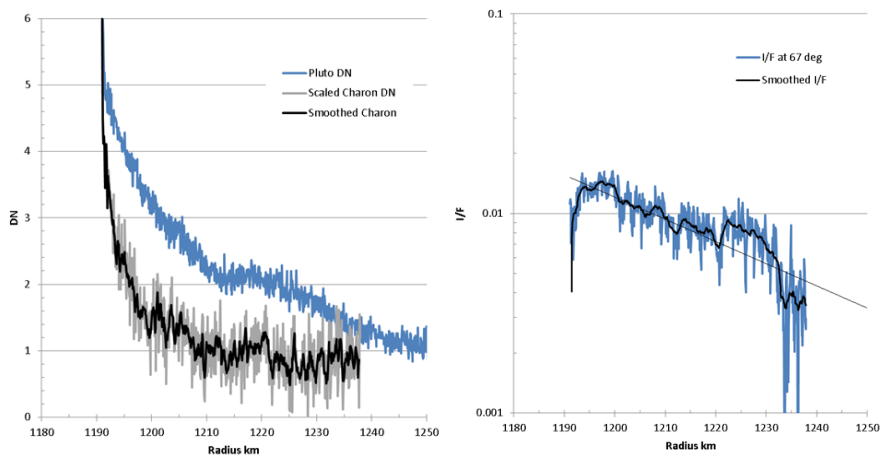

Figure 15. (left) DN profiles from Pluto at 67° phase angle; (image MET0299179658) in blue and Charon (image MET0299180406) in gray. Charon profile smoothed over 20 samples in black. (right) I/F from excess of Pluto DN profile over scaled Charon profile, with exponential fit shown indicating 39 km scale height.

**Haze at 148° phase** The sequence P_HIPHASE_HIRES (Table 1) just after closest approach observed Pluto at 93 m/px resolution and at 148° phase angle. The image MET0299181359 from P_HIPHASE_HIRES is shown in Fig. 5. The image was rotated and a column-averaged, row profile of brightness was extracted within the box shown to average out image row striping. At 148° phase angle, the haze above the limb and the sunlit surface of Pluto are seen to have similar brightness (left panel, Fig. 16). At the lower phase angles 20° and 67°, the sunlit surface is much brighter than the haze. At higher phase angles (e.g., Fig. 6 at 169° phase), the haze is brighter than the sunlit surface of Pluto.

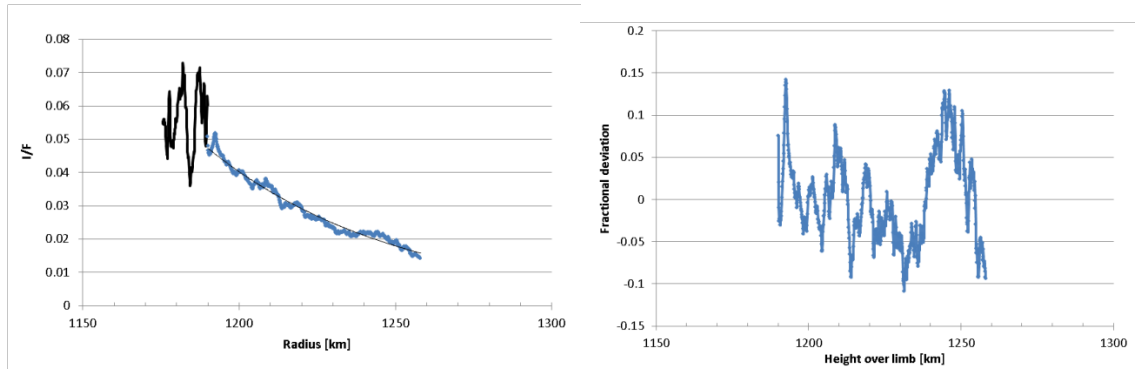

**Figure 16** Haze I/F profiles at 148° phase angle. (left) I/F of haze above the limb in blue, and sunlit surface below the limb in black, showing haze and sunlit surface at similar brightness. (right) detrended haze brightness using exponential fit; fractional deviation is [(I/F)/trend]-1.

**Table 4. Phase Functions of Pluto Haze Particles**

| Phase Angle | Peak I/F of Haze | I/F at 45 km |
|---|---|---|
| 20° | 0.02 | 0.004 |
| 67° | 0.014 | 0.004 |
| 148° | 0.05 | 0.02 |
| 167° | 0.3 | 0.15 |

**Phase Function.** The phase function of Pluto's atmospheric haze at the visible wavelengths observed by LORRI (pivot wavelength 607.6nm) is compiled in Table 4. The Pluto observations were acquired during a 19 hour time span and over various Pluto locations in the northern hemisphere, and effects of temporal or spatial variations may not be fully accounted for. The two columns of Table 4 show phase functions compiled at the peak I/F just above the surface and at an altitude 45 km (~one scale height above the surface).

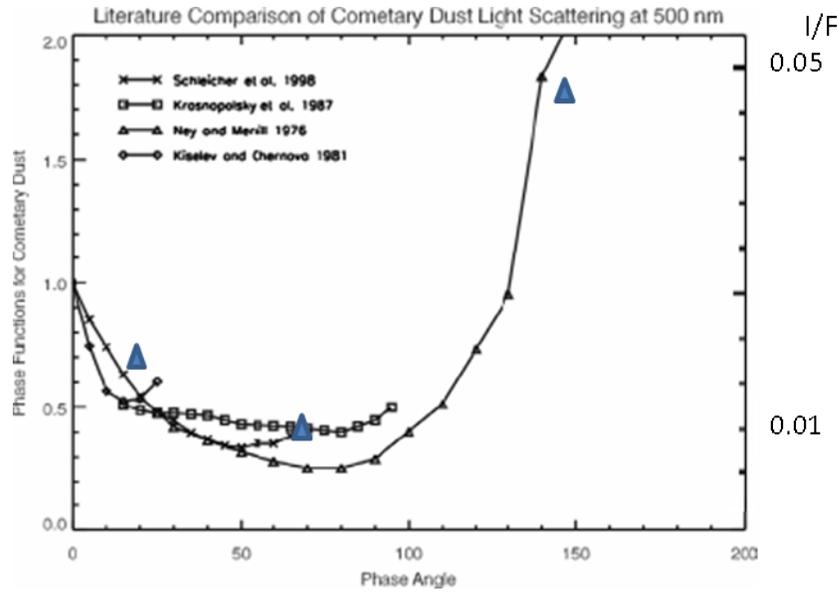

**Figure 17 Pluto haze phase function (triangles) compared with cometary dust, linear scale**

A notable difference is found between the phase function just above the surface near the peak I/F and that at 45 km altitude. Namely, the former displays a backscatter lobe which is absent in the latter. This difference is reflected in the low scale heights for haze reported in the backscatter geometries of Figs. 14 and 15, causing the backscatter lobe found at the lower altitude to disappear at the higher altitude.

Pluto's haze is strongly forward scattering in the visible, becoming brighter than the sunlit surface of Pluto at large phase angles above about 148°. Pluto's forward scattering phase function is similar to that of comet dust (see Fig. 17; Schleicher et al. 1998, Krasnopolsky et al. 1997, Ney and Merrill, 1976, Kiesler and Chernova 1981) and of Saturn's G-ring dust (see Fig. 18; Hedman and Stark 2015). In the phase function near the peak I/F, there is a strong, narrow forward scattering lobe and a back scatter lobe, as seen from comet dust and as expected for Mie scattering by spherical particles. However, the phase function at the higher altitude, with a strong forward scattering peak but without a backscatter lobe, is more similar to the phase functions found for haze in Titan's atmosphere (Lavvas et al. 2010).

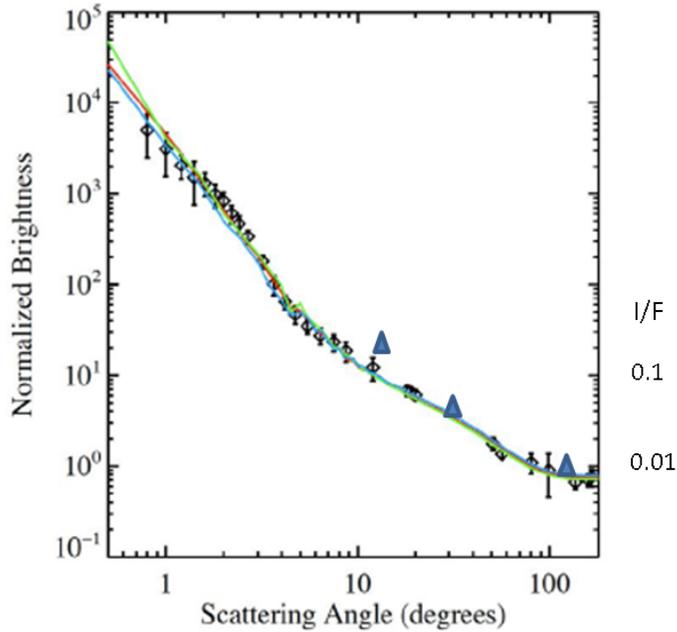

**Figure 18** Pluto haze phase function (triangles) compared with Saturn's G ring dust, log scale. G-ring data and Mie scattering fits from Hedman and Stark, 2015

5. **Haze Physical Properties**

The optically thin haze in Pluto's atmosphere is observed to altitudes above the surface >200 km with haze scale height of typically 50 km (Gladstone et al. 2016). The peak I/F~ 0.3 just above the surface at 167° phase angle as seen both by LORRI and by MVIC in its red filter, whose pass band 540-700 nm is centered close to the LORRI pivot wavelength of 607.6 nm. The observation sequence was PC_MULTI_DEP_LONG_2. The MVIC color images indicated a blue color (Gladstone et al. 2016) for Pluto's haze, with peak *I/F* reaching 0.7 to 0.8 in the blue filter (400 to 550 nm). Although the blue haze color is consistent with Rayleigh scattering by small (radii r~0.010 µm) particles, the strongly forward scattering visible phase function suggests much larger particles (r > 0.1 µm). Gladstone et al. (2016) suggest that aggregate particles (randomly shaped aggregates of ~0.010-µm spheres could satisfy both constraints. The MVIC blue/red color ratio increases with altitude, consistent with smaller particles at higher altitudes.

Physical properties of the Pluto haze particles are estimated following Gladstone et al. (2016). For tholin-like particles with optical constants of n =1.69 and k =0.018 (real and imaginary parts of complex refractive index), Mie scattering by particles of at least 0.2 µm radius would yield $P(165°) \sim 5$. From (1), the line-of-sight scattering optical depth would be ~0.24, and the vertical optical depth would be ~0.018. For particles of 0.2 μm radius, Mie theory gives $\sigma_s = 3.4 \times 10^{-9}$ cm$^2$ with $Q_s$=2.7, and the number density of haze particles becomes $n = 1.2$ cm$^{-3}$ for a haze scale height of 50 km. These estimates may be compared to those inferred for the detached haze layer at Titan ~520 km above Titan's surface (Lavvas et al. 2009), namely, an apparent radius of ~0.040 μm and number density of ~30 particles cm$^{-3}$.

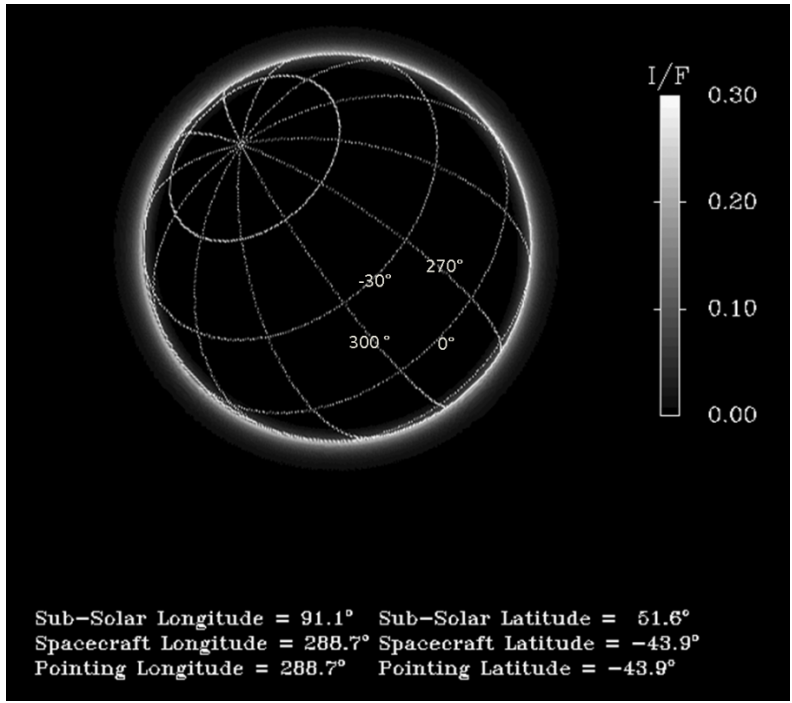

**Figure 19. Haze illumination model for geometry of Fig 2. Scale height of 50 km and azimuthally uniform distribution around the limb are assumed. Phase angle is 165.9°; longitude 300° and latitude -30° indicated. South Pole faces viewer. Model does not reproduce brightness enhancement to north in Fig. 2, indicating a greater haze column in the north.**

Fig. 19 shows a haze brightness model which calculates the solar illumination over the dayside versus night side limb. This is a single scattering model, assuming that the spatial distribution of haze is azimuthally uniform around the limb. The illumination, viewing geometries and image orientation are those of Fig. 2. The sub-solar longitude and latitude on Pluto, and the sub-spacecraft longitude and latitude, are given in Fig. 19. Over sunlit limb, haze particles in the line-of-sight are illuminated by the sun, but over the night side, lines of sight pass into Pluto's shadow where haze particles are not illuminated. However, the model of this illumination effect is approximately 20% brighter in the north than over the equator. The strong brightness asymmetry seen in Fig. 2, with haze brighter in the north, is not reproduced. The greater brightness of Pluto's haze in the north compared to the equator, by a factor of 2 to 3 according to Fig. 11, is not explained by illumination geometry but indicates a greater optical depth of haze particles in the north.

Using the I/F profiles retrieved at different phase angles and under the assumption of single scattering conditions we can retrieve further constraints on the properties of the haze particles. Assuming a constant particle size and a density of particles that decreases exponentially following a scale height, we can model I/F profiles based on Eq. (1) assuming different particle shapes and refractive indexes. Such simplified models provide useful information for particle properties near the peak I/F above the surface. Figs. 20 and 21 describe two such models, for spherical particles and for aggregate particles, respectively.

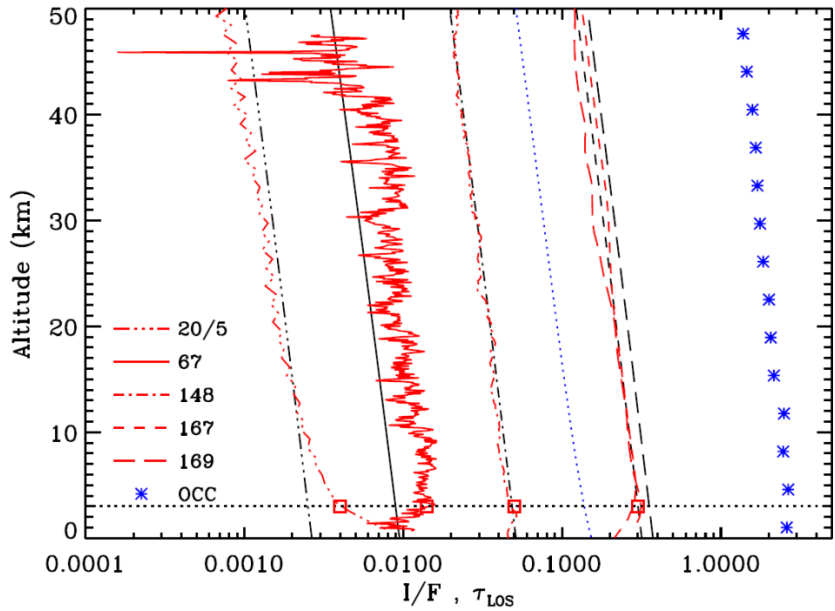

**Figure 20. I/F profiles modeled at five phase angles, assuming spherical particles (Mie scattering). Red curves are LORRI I/F profiles (section 4) and blue asterisks are extinction profiles from the ALICE UV solar occultation. Simulated I/F profiles (black curves) match the visible I/F profiles versus phase angle, but the UV extinction (blue dotted line) is too low. Observed and simulated I/F profiles at 20° phase are divided by a factor of 5. Red squares are I/F values of Table 4.**

For spherical particles, the peak I/F values reported in Table 4 at four different phase angles at visible wavelengths can be reproduced with spheres of 0.5 μm radius and a near surface haze density of 0.1 cm$^{-3}$ that decreases with altitude following a scale height of 50 km (Fig. 20). A log-normal size distribution for the particles is assumed with a standard deviation of 0.2, and the refractive index of the particles is similar to that of Titan's haze analog tholins (Khare et al. 1984). These calculations include the contribution of scattered light from Pluto's surface that can enhance the I/F relative to the values obtained from the direct scattering of solar flux, for medium phase angles (particularly for the 67° phase angle in our data set). We calculate this contribution assuming a surface albedo of 50%. The simulated I/F profiles of Fig. 20 agree with the visible observations. However, the resulting line-of-sight extinction opacity at UV wavelengths is more than a factor of 10 smaller than that retrieved from the ALICE solar occultation.

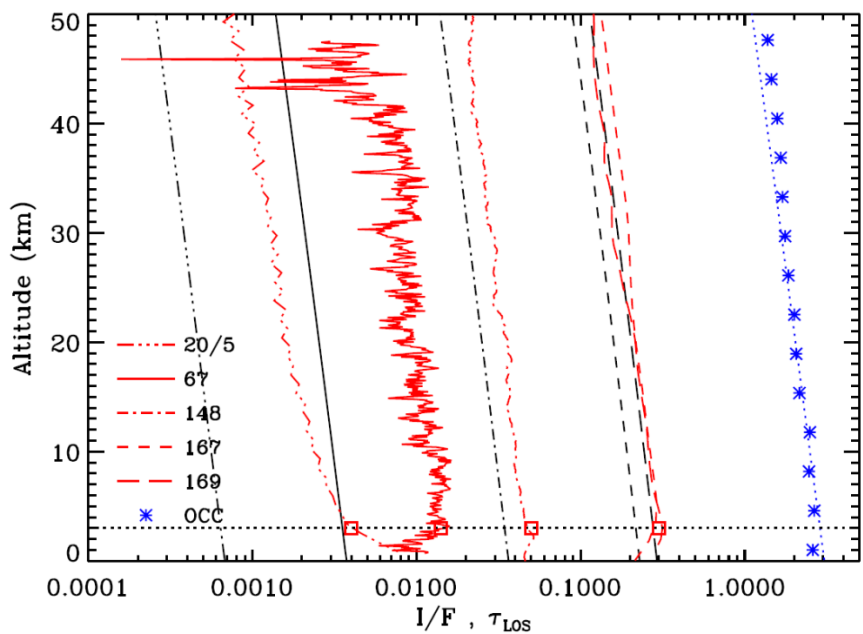

**Figure 21. Same as in Fig. 20, but assuming aggregate particles with fractal dimension of 2. Primary particle radius is set to 0.01 μm and the bulk radius at 0.15 μm. Aggregate particles give UV line-of-sight opacity consistent with the solar occultation, but under-estimate the visible back scattering at 20º and 67º phase angle.**

Aggregate particles can induce a significantly higher opacity at UV wavelengths as is well understood from studies in Titan's atmosphere. Considering aggregate particles composed of primary spherical particles with 0.01 μm radius and a particle shape corresponding to a fractal dimension of 2 (like those in Titan's main haze layer, Tomasko et al. 2008) generates a UV line-of-sight extinction opacity consistent with that from the solar occultation, while also providing strong forward scattering that reproduces the observed visible I/F at 148º and 166º phase angle (Fig. 21). In this case the particle bulk radius is 0.15 μm and the particle density close to the surface is 15 cm$^{-3}$ (the particle scale height and refractive index are kept the same as for the spherical particle case). However, the resulting back scattering for aggregate particles gives a too small backscatter lobe by factor ~3. Changes in the primary particle size and the other free parameters (bulk size, density, and refractive index) can increase the back scattering contribution, but at the expense of the fit quality at forward scattering.

However, the Titan-like aggregate particle model in Fig. 21 would be consistent with both the UV extinction from the solar occultation and the higher altitude visible phase function of Table 4, in which there is a strong forward scattering peak without a back scattering peak. The different phase functions given in Table 4 near peak I/F and at 45 km altitude indicate different physical properties for haze particles just above the surface (lowest 15 km) versus at higher altitude. A 0.5 µm spherical particle model (Fig. 20) reproduces the visible phase function of haze just above the surface, whereas an aggregate particle model (Fig. 21) matches the higher altitude visible phase function and the UV extinction. The Mie scattering model does not match the UV extinction, but the UV solar occultation is sensitive mainly to much higher altitudes.

## 6. Haze Production Mechanisms

The atmospheric haze observed by New Horizons is quite different than that seen on Triton, or that expected for Pluto based upon the photochemical/vertical transport condensation model (Summers et al., 1997). There is a low altitude haze layer at ~5-15 km altitude above Pluto's surface that appears to correspond to the expected photochemical/condensation haze layer. Given that the New Horizons Alice experiment observed both $C_2H_2$ and $C_2H_4$ in abundances comparable to those predicted by photochemical models, and as noted above the mixing of these species downward would produce highly supersaturated conditions below about 20 km altitude, the most plausible explanation for the haze layer at ~5-15 km altitude is that it is produced by direct condensation of photochemical products, if sufficient nuclei are present.

### Nucleation in Pluto's Atmosphere

For a nucleation rate equal to the inferred particle production rate of ~1 $cm^{-2}s^{-1}$ as suggested in Gladstone et al. (2016), there are several viable nucleation sources in Pluto's atmosphere. Near-surface horizontal winds may mobilize surface particles if wind velocities are sufficiently high and if surface cohesion is low. Lofting of surface particles may cause the observed dark streaks on Triton (Sagan and Chyba, 1990), and the equatorial "dunes" on Titan (Lorenz, et. al., 2006).

Adopting the Sagan and Chyba (1990) formulation for eolian mobilization of particles on Triton, we find that lofting of particles larger than 0.001 µm from a cohesionless surface can occur on Pluto if the horizontal wind velocity is > 1 cm $s^{-1}$. However, even extremely small degrees of surface cohesion can reduce lofting by many orders of magnitude. The REX observations of a near surface boundary layer on the entry occultation above Sputnik Planum (Gladstone et. al., 2016) with a vertical extent of less than 4 km would suggest that any small pickup particle that were lofted from the surface would be constrained to those altitudes (i.e., to the level at which mixing can transport the particles). The REX exit occultation showed no boundary layer, so if lofting of surface particles does occur on Pluto it may be limited in geographical extent. Even if the pickup occurs only in the lowest few kilometers of Pluto's atmosphere, that might be sufficient to provide condensation nuclei for a very low altitude fog.

Another possible source of condensation nuclei is solar EUV and cosmic ray ionization. In contrast to eolian lofting of particles, cosmic ray ionization occurs throughout the vertical extent of Pluto's atmosphere (Summers et al., 1997; Krasnopolsky and Cruikshank, 1999). The estimated total peak ionization rate below ~1000 km altitude is ~ $2 \times 10^{-4}$ cm$^{-3}$s$^{-1}$ for a peak electron density there of 14 cm$^{-3}$. For an atmosphere with a scale height of 50 km, the column ionization rate is ~$10^3$ cm$^{-2}$s$^{-1}$. The portion of ions facilitating the formation of condensation nuclei is unknown, but if even only 1% do then there is a sufficient nucleation rate to produce ~1 cm$^{-3}$ s$^{-1}$ through Pluto's atmosphere.

Incoming Kuiper Belt dust on Pluto (Poppe, 2015) of ~50 kg per day, that is presumed to be composed mostly of water ice, could contribute about 0.5 cm of surface water ice to Pluto's surface over the age of the solar system. If the incoming dust particles are made entirely of water ice, and are ablated in the uppermost atmosphere, then the downward transport of water molecules would be ~ $1 \times 10^5$ cm$^{-2}$s$^{-1}$. If each water molecule acts as a nucleation site, then that would be a vast number of condensation nuclei.

Overall, it appears that solar EUV and cosmic ray ionization, along with incoming ablated cosmic dust particles, can readily produce sufficient condensation nuclei throughout Pluto's atmosphere to explain haze formation in supersaturated regions. The haze that extends to altitudes ~200 km is surprising because the hydrocarbons are sub-saturated there, and direct condensation should not occur (Fray and Schmitt, 2009). The atmospheric temperature peaks above 110K near 30 km altitude, falling to 76K at 200 km (Gladstone et al. 2016). It may be possible that above 200 km altitude an atmospheric temperature decrease may allow super-saturation of certain species, in particular HCN (see below).

### Similarity with Titan's Detached Haze Layer

Neither the embedded bright and narrow haze layers, nor the background haze extending to above 200 km altitude, was expected prior to the New Horizons encounter. Titan also has a mainly $N_2$ atmosphere with upper atmospheric hazes similar to those observed on Pluto. At high altitude Titan has a "detached haze layer" (Lavvas, et al., 2009), which is separated from the broad lower atmospheric haze and which varies in altitude from ~500 to ~350 km with season around equinox. This detached haze layer occurs at altitudes where the temperature is too high for direct condensation of the hydrocarbons and nitriles known to be present, much the same situation as on Pluto. Titan's detached haze layer also occurs in the 1-10 microbar level, i.e., overlapping the pressure level where extended haze is observed on Pluto (see Fig 22).

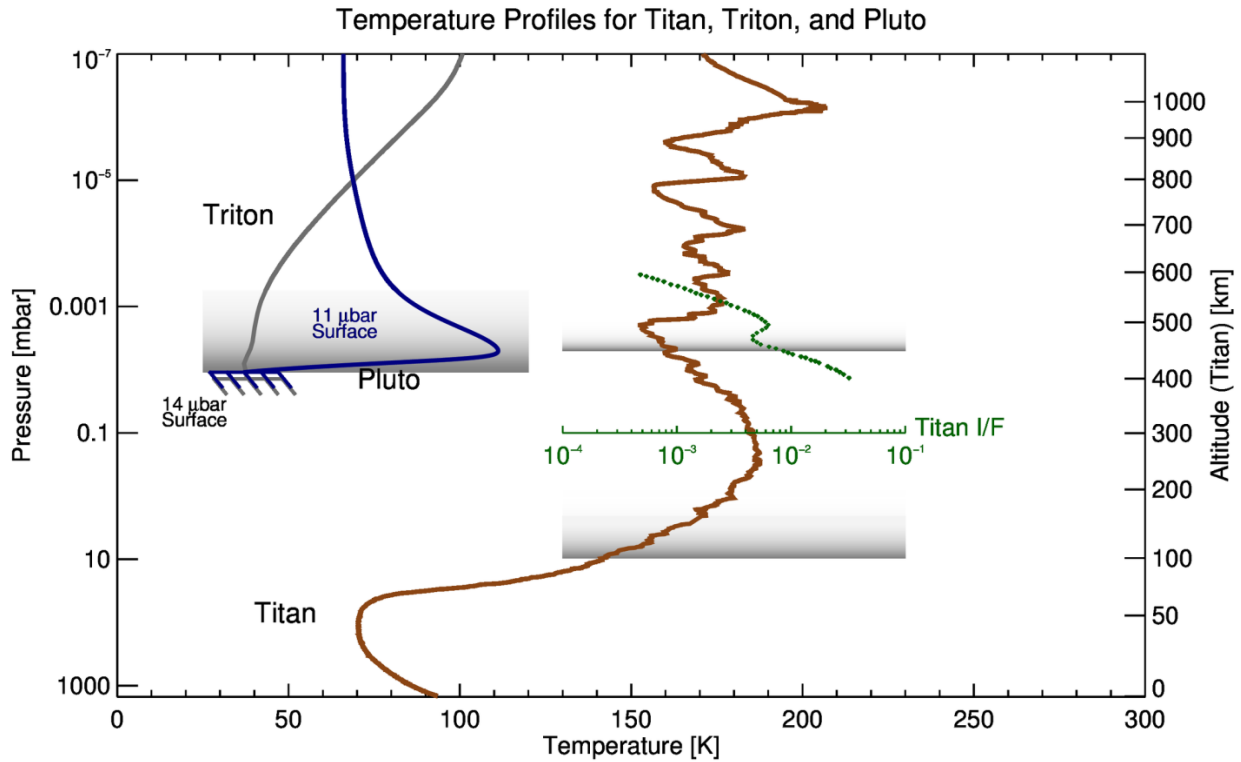

**Figure 22. Titan's atmospheric temperature from the Cassini Huygens probe is shown versus temperature and pressure for Titan's atmosphere. The blue (black) curve is Pluto's (Triton's) temperature plotted on the pressure scale. Titan's I/F for altitude range ~400-600 km above Titan's surface is shown in green. The shaded region above Pluto's surface shows the lower ~200 km of Pluto's atmosphere where haze is observed.**

Lavvas et al. (2010) have shown that Titan's haze can be understood as forming in Titan's ionosphere at ~1000-1200 km altitude. These processes involve low energy electron collisions with nitriles (HCN and $HC_3N$) initiating the formation of negative ions $CN^-$ and $C_3N^-$ and eventually leading to very large negatively charged macromolecules that attract positive ions (Vuitton et al.,2009), forming aerosols that grow, settle downwards and eventually coagulate collisionally into aggregates. The key point is that ionospheric chemistry occurs in the presence of hydrocarbons on Titan and Pluto, but not Triton as methane is destroyed in the lower atmosphere. These aggregates will then grow through chemistry and diffusion to form aerosols of thousands of AMU. These aerosols settle downwards as they grow, and at some point coagulation begins to dominate growth, resulting in formation of fractals. The detached layer occurs near where this transition occurs.

The REX experiment on New Horizons has not reported a detection an ionosphere on Pluto based on analysis to date. Bagenal et al. (2016) estimated with density, temperature, and composition measurements from New Horizons (Gladstone et al., 2016), a peak electron density of $\leq$1300 cm$^{-3}$ at an altitude of ~700 km. This peak electron density may be below the REX sensitivity level and occurs about 500 km above the top of the observed haze, where processes similar to those above Titan's detached haze layer could occur.

We, expect an overall reduction by a factor of ~12 in the ionization rate in Pluto's atmosphere below that at Titan in part because of Titan's larger distance from the Sun, and because Pluto's atmosphere is not exposed to ionization from charged particles in Saturn's magnetosphere. A nucleation mechanism for haze particles operating in Pluto's atmosphere, such as that which exists in Titan's haze, might explain why Pluto has a haze in the hottest part of its atmosphere where the hydrocarbons are sub-saturated.

### Haze Particle Sedimentation

Ultimately all of the particulates in Pluto's atmosphere are lost to the surface. The sedimentation rate downward through the atmosphere depends upon their size, shape, density, and gravity, as well as the characteristics of the atmosphere itself. From the ionosphere down to the observed haze layer at ~200 km altitude, the monomers produced by the Titan mechanism will settle downward and grow by collision of hydrocarbon molecules. Their relative motion is due to Brownian motion at high altitudes, and by simple terminal fall velocity at lower altitudes. We ignore the effects of electric fields and assume that the vertical motion of particles is controlled by gravity and friction (viscosity) with the background atmosphere.

Figure 23 compares several time scales relevant to the distribution and growth of haze particles, for several particle sizes: 0.1, 0.2, and 0.5 μm. As noted above, the inferred size depends upon observing geometry and wavelength, and at this point we do not have a unique solution for particle size that will explain all New Horizons observations. This is almost certainly due to the fact that the observed haze represents a range of sizes as well as a mix of fractal particles. For Titan fractals range from D=2 to D=3 with a complex altitude dependence (Lavvas et al., 2010), but here we show results for Pluto with only D=2.

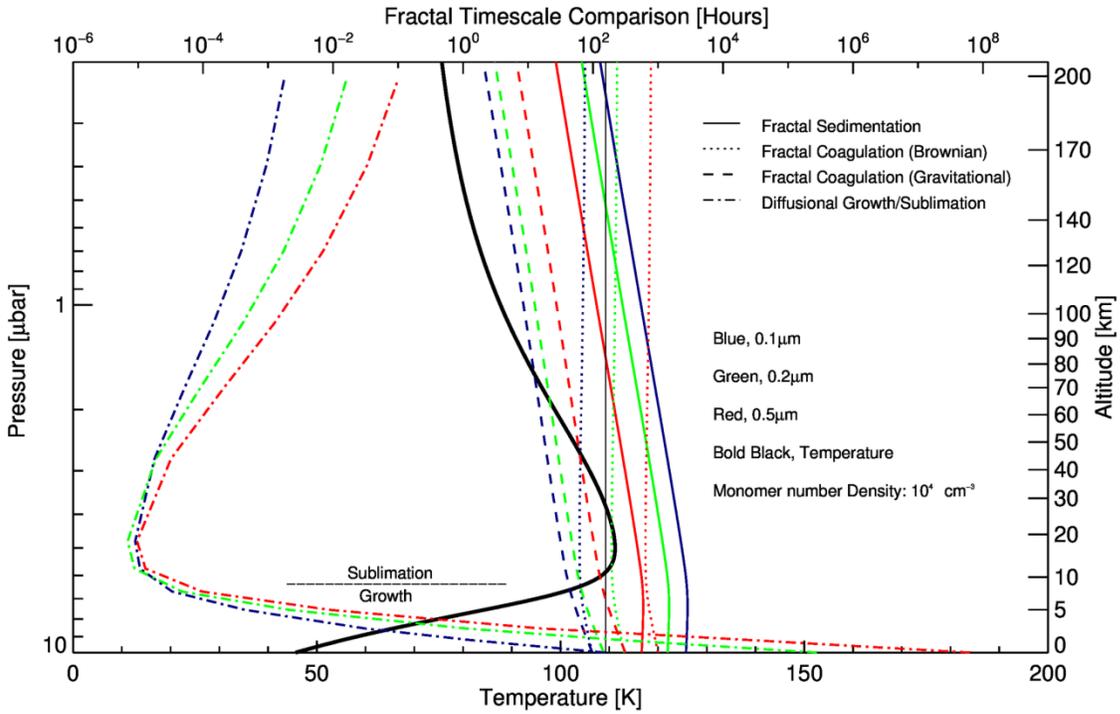
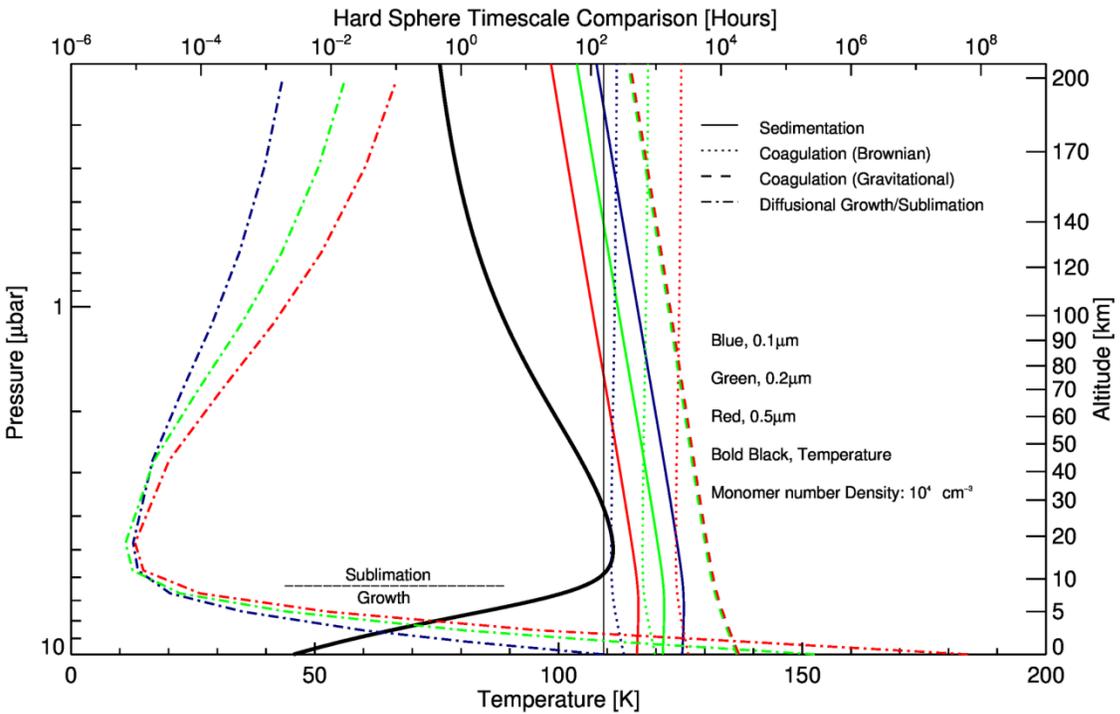

Figure 23. Timescales for sedimentation, coagulation, and diffusion are defined, respectively, as the time necessary for the primary particle to fall one scale height, grow to double its original size through collisions with monomers, and change size by factor of 2 due to diffusion of a hydrocarbon or nitrile gas molecule to or from the primary particle. The primary particle is assumed to be in a vertically constant background of monomers of size 0.05 μm, taken from Lavvas et. al. (2010). The vertical black line denotes a Pluto day of 6.38 Earth days; the bold black curve is the temperature profile.

The sedimentation timescales in Fig. 23 are slightly different for hard spheres and fractal aggregates. The correction factor for the porosity of the fractal aggregate has been taken from Lavvas et al. 2010. The approximately 30% adjustment in timescale can be seen for each particle size due to the increase in drag that occurs with D=2 fractals that have a larger drag than their corresponding equivalent size hard spheres. Coagulation timescale for hard spheres is shown for both Brownian motion and gravitational settling.

The coagulation kernel due to Brownian motion is again taken from Lavvas et al. (2010) and is caused by the statistical bombardment of monomers with the primary particle. The gravitational coagulation Kernel is taken from Seinfeld and Pandis (1998), and is caused by the bombardment of monomers with the primary particle due to a difference in the sedimentation velocity. The expression for particle size, dependent on time, is found by integrating the equation for rate of collisions due to each process, while assuming the rate doesn't change considerably over the doubling of the particle size.

The density of monomers formed in Pluto's ionosphere is unknown at present. A constant number density of monomers for coagulation is assumed to be $10^4 \text{cm}^{-3}$, taken from the approximate temperature—pressure regime coinciding with both Titan and Pluto (Lavvas et al., 2011, Figure 4). A collision efficiency factor of unity is also assumed.

The diffusional growth timescale is calculated from a density profile of $C_2H_2$ from the Caltech KINETICS photochemical model (Wong et al., 2015). The diffusional growth is driven by the difference (gradient) in partial pressure of $C_2H_2$ in the ambient atmosphere compared to the saturation pressure at the surface of the particle. The rate of change equation for the radius of the primary particle is integrated assuming the saturation vapor pressure at the surface of the particle (from the Clausius-Clapeyron equation) does not change considerably over the doubling of size. The saturation vapor pressure for $C_2H_2$ far from the particle for the low temperature-pressure regime seen in Pluto's atmosphere is found by an empirical formula with coefficients taken from N. Fray and B. Schmitt (2009).

The sedimentation time scale of a haze particle is the vertical sedimentation velocity of the particle divided by the atmospheric scale height. The sedimentation time for hard spheres, using 0.2 μm particles as an example, ranges from ~60 hours at the top of the haze layer at 200 km altitude, to ~500 hours near the surface, illustrating how the fall velocity is rapid in the thinner upper atmosphere. The sedimentation time is about ~1.3 times smaller for hard spheres than that for fractals with the same mass but dimension D=2.

The coagulation time scale is defined as the time it takes coagulation to double the mass of the collector particle. Thus it depends sensitively upon both the relative velocity of the particles and the mass of the collector. For larger collector particles, the time scale begins to decrease rapidly. Thus we see that the largest particles of 0.5 μm take extremely long times to grow by Brownian motion in the lower atmosphere.

The diffusion growth time scale is interesting in the lower atmosphere, including what may be termed a photochemical haze layer below ~20 km altitude. The monomers are initially spherical at high altitudes ~1000 km, become fractals as they fall into the haze layer at ~200 km, and continue to grow as fractals until they fall out of the main haze layer into the lower temperature region below ~20 km altitude. Here the diffusion time scale for of supersaturated hydrocarbons and nitriles is extremely short, leading to sublimation of these species above ~10 km and growth by condensation below ~10 km. The diffusion timescales increase rapidly in the bottom ~2 km, approaching and exceeding sedimentation timescales. Thus we expect rapid "coating" of the fractal particles, which may become spherical before they settle onto the surface.

### 7. Gravity Waves and Haze Layering

In Gladstone et al. (2016) gravity waves, as a result of orographic forcing, were advanced as a plausible explanation of the distinct haze layers in Pluto's atmosphere detected by New Horizons imagery. Solar induced sublimation of surface ices (principally $N_2$, with minor amounts of $CH_4$ and CO ices) drives transport to colder, lower surface regions, while buffering atmospheric pressure and composition. Subsequent condensation constrains pressure variations in the atmosphere above the first ½- scale height to $\Delta p/p <$ 0.002 for a surface pressure ~10 μbar (Young, 2012). With very little pressure variation in the current atmosphere on a global scale, horizontal winds are expected to be weak (i.e., no more than about 10 m s$^{-1}$, e. g. Forget et al., 2016). In spite of weak winds, they are still capable of generating internal gravity (buoyancy) waves driven by orographic forcing (wind blowing over topography). Mountains / mountain ranges with heights of 2-3 km have been detected by New Horizons imagery (Moore, et al., 2016).

The gravity wave hypothesis relies on the fact that weak winds blowing over topography can reach saturation amplitudes close to the surface. By saturation we mean the amplitude limit whereby the buoyancy restoring force vanishes. This occurs where the sum of the wave and mean temperature gradients render the atmosphere adiabatic and the vertical parcel/wave velocity, $w'$, is equal to $w_g$, the vertical group velocity defined below. At saturation, haze particle displacements, $\zeta$, reach their maximum amplitude given by, $\zeta_{sat} \simeq \lambda_z / 2\pi$, where $\lambda_z$ is the vertical wavelength of the gravity wave. The layering parameter introduced in Gladstone et al. (2016), $\eta = (0.5\lambda_z + 2\zeta)/(0.5\lambda_z - 2\zeta)$, thus takes on maximum and minimum values of 4.5 and 0.22.

The gravity wave fields are represented by complex amplitudes $\Psi' = \psi(z)\exp[i(kx+ly-kct)]$, where the x axis is the "zonal" prograde direction with the zonally forcing wind blowing over orography in the positive x direction, the positive y direction of this Cartesian coordinate system is toward the north pole, and the angular frequency as seen by an observer on Pluto is kc, with c the horizontal zonal phase speed. The horizontal wavenumbers $k$ and $l$ for the $x$ and $y$ directions have corresponding horizontal wavelengths $\lambda_x$ and $\lambda_y$. Gladstone et al. (2016) anticipated without direct observational evidence that the gravity waves were standing waves, and c was set equal to zero (c = 0). Pluto's rotation will not play an important role in wave properties and the coordinate system can be rotated to have the positive $x$ axis aligned with the positive direction of the mean winds, $u_0$. The orographic forcing of gravity waves is specified by $w'(z=0) = w_0 = u_0(z)\partial h/\partial x$, where $h = h_0 \cos(kx)$ and $h_0$ is the Fourier height of the topography.

The gravity waves are treated as propagating in a mean wind with vertical shear and the relevant equations may be found in Lindzen (1990), but here $z$ is altitude instead of $-\ln(p/p_s)$, where $p_s$ is the surface pressure. The equations are reduced to a single, second order differential equation for the vertical structure amplitude of $w'$, $W(z)$.

(2) $\quad \dfrac{d^2W}{dz^2} + m^2(z)W(z) = 0$ ; where $m(z) = \left[\dfrac{N^2-\omega^2}{\omega^2-f^2}(k\delta)^2 - \dfrac{1}{4H^2} + \left(\dfrac{1}{H}\dfrac{du_0}{dz} + \dfrac{d^2u_0}{dz^2}\right)\dfrac{1}{c-u_0}\right]^{0.5}$

Here $m(z)$ is the vertical wavenumber, defined from the gravity wave dispersion relation, and $H = H(z)$ is the density scale height, $f$ is the Coriolis parameter, $\delta = [1+(l/k)^2]^{0.5}$ is the aspect ratio, and $\omega^2 = [k(c-u_0)]^2$ is the square of the Doppler shifted angular frequency measured by an observer riding with the mean wind. The sign of $\omega$ is chosen to ensure that the vertical group velocity is directed upward and the vertical phase velocity is directed downward. One now assumes that all quantities in $m(z)$ are slowly varying functions of altitude and that the solution to Eq. (2) can be approximated by the WKBJ short wavelength approximation solution as $(mH)^2 \gg 1$:

$$(3) \quad w' = w_0 \left(\frac{\rho(0)}{\rho(z)}\right)^{0.5} \left(\frac{m(0)}{m(z)}\right)^{0.5} \exp[i(kx+ly-kct)]\exp\left[-i\int_0^z m(z)dz\right]$$

To integrate this solution from the surface to any altitude the integral is converted to a summation with the atmosphere divided into vertical step sizes of 0.25 km. At each step in altitude the amplitude of w' is compared with its saturation amplitude, $w_g$, which is calculated by

$$(4) \quad w_g(z) = \frac{\partial \omega(z,k,m)}{\partial m} = \frac{-k\delta\, N(z)\, m(z)}{\left[m^2(z) + \frac{1}{4H^2} + (k\delta)^2 - \left(\frac{1}{H}\frac{du_0}{dz} + \frac{d^2u_0}{dz^2}\right)\frac{1}{c-u_0}\right]^{1.5}}$$

The sign of $m(z)$ is chosen to ensure that the vertical group velocity is directed upward and the vertical phase velocity is directed downward. Thus $m(z)$ must be < 0. If $|w'(z)| > w_g(z)$, w'(z) is replaced by $w_g(z)$ and the integration proceeds to the next level.

Exploration of parameter space leads to a convergence of constraints that yield the desired solutions and plausibly explain the haze layering by gravity waves. They are: 1) $\omega^2 \ll N^2$, which implies the gravity waves are hydrostatic, and 2) $\omega^2 \gg f^2$, which implies that the gravity waves are not inertia gravity waves and influenced by rotation and that energy is not mostly transported horizontally, for small values of l, (e.g., meridional wavelength $\lambda_y \sim$ 3600 km). Also $\lambda_x$ must be less than 360 km, and wave saturation is possible over an extended region of the atmosphere.

The corresponding solution for T'(z) is

$$(5) \quad T'(z) = i\frac{T_0(z)N^2}{g\,k\,u_0} w'(z)$$

where $T_0$ is the mean temperature and g is the gravitational acceleration. The vertical wavelength is calculated from the gravity wave dispersion relation at each altitude, which is a function of the background atmosphere properties (Fig. 24), and the horizontal wavelengths (wavenumbers). The particle displacement is calculated from the Lagrangian equation evaluated riding with the mean wind,

(6) $$\frac{D\zeta}{Dt} = \left(\frac{\partial}{\partial t} + u_0(z)\frac{\partial}{\partial x}\right)\zeta = ik[u_0(z)-c]\zeta \equiv iku_0(z)\zeta = w'$$

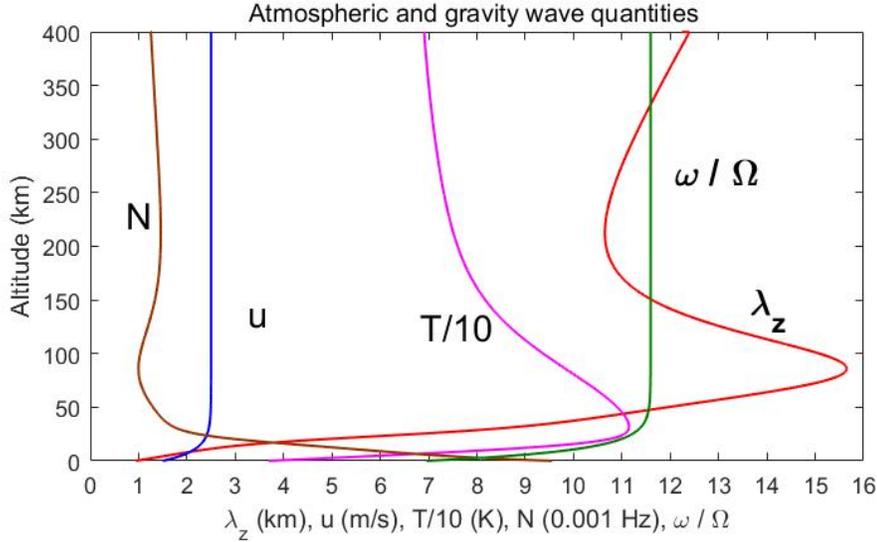

**Figure 24.** Red line is vertical wavelength in km, blue line is the zonal wind velocity, $u_0$, in m s$^{-1}$, mean temperature profile divided by 10 is the magenta line, brown line is the buoyancy frequency in units of mHz, and green line is the angular Doppler-shifted frequency, $\omega$, in units of the Pluto's rotation rate. Calculation assuming $\lambda_X \sim 120$ km (zonal wavenumber $k = 2\pi / \lambda_X \sim 60 / R_P$), meridional wavelength $\lambda_y \sim 3600$ km

The prime influence of zonal winds, $u_0$, is on the vertical wavelength $\lambda_z \sim 2\pi u_0 / N\delta$ to lowest order from the dispersion relation in Eq. (2). The vertical wavelength is also impacted by the ratio of the horizontal wavenumbers via $\delta$ and thus the spacing of the haze layers is mostly determined by $u_0$ and $\delta$, because $T_0(z)$ and $N(z)$ are constrained by Alice and REX occultation data.

Any surface vertical forcing velocity $w_0 \geq 0.004$ m s$^{-1}$ yields solutions to the gravity wave equation that reach saturated amplitudes by altitude 4 km where the wave temperature of $T' \sim 0.6$ K, due to the small adiabatic lapse rate (~0.62 K km$^{-1}$ at the surface, decreasing with altitude), and where the vertical parcel velocity of $w' = w_g \sim 0.006$ m s$^{-1}$, the vertical group velocity. For the threshold value of $w_0 = 0.004$ m s$^{-1}$ to achieve saturation amplitudes at 4 km, one only needs the product of $u_0 h_0 = 7.5$ with $u_0$ in cm s$^{-1}$ and $h_0$ in km, for example, $u_0 = 5$ cm s$^{-1}$ and $h_0 = 1.5$ km (or 100 cm s$^{-1}$ and 0.075 km) blowing over topography characterized by horizontal wavelength $\lambda_X \sim 120$ km (zonal wavenumber $k = 2\pi / \lambda_X \sim 60 / R_P$), meridional wavelength $\lambda_y \sim 3600$ km, our preferred values. Finally the wave temperature amplitudes are sufficiently small in comparison to the mean temperature profile to not interfere with the retrieval of temperature from the Alice solar occultation and the REX radio occultation.

The saturation amplitude of the parcel displacement is calculated to lowest order from Eqs. (6, 4, and 2)

(7) $$\zeta_{sat} = \left|\frac{w_g(z)}{i k u_0(z)}\right| \approx \frac{Nk\delta m(z)}{k u_0(z) m^3(z)} = \frac{N\delta}{u_0(z) m^2} = \frac{1}{|m|} \ ; \quad \text{where} \ m^{-1} \approx \frac{N\delta}{u_0(z)}$$

Now the hypothesis is to displace haze particles by the gravity wave vertical velocity to compress and dilute haze particles in any given volume. In Fig. 25 the vertical phase/particle velocity, $w_j$ (magenta line)

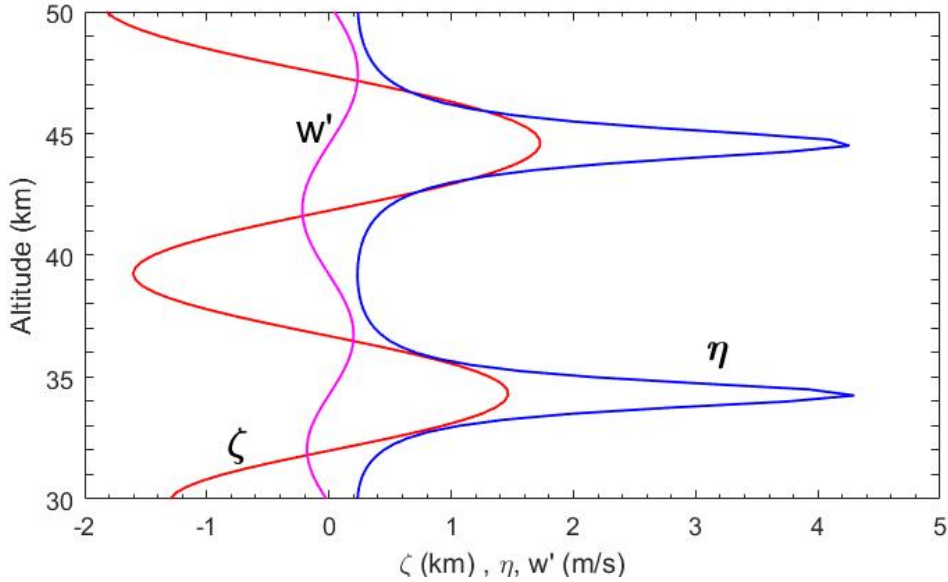

Figure 25 The red line is the parcel displacement in km, the magenta line is vertical phase/particle velocity w' in m/s, and the blue line is the non-dimensional layering parameter η

shows upward motion from z = 35 to 39.25 km and downward motion from 39.25 to 44.5 km. This leads to a convergence and compaction of haze particles in the 38-40.5 km region, where the particle displacement, $\zeta$ and red line, has a maximum negative value. Also the layering parameter, $\eta$ and blue line, has a minimum value there. Thus minimum values of η correspond to maximum concentrations of haze particles via compression and maximum values of η correspond to minimum concentrations of haze particles via rarefaction.

To calculate the density changes induced by gravity waves, we include horizontal compression and rarefaction which approximately cancels vertical compression and rarefaction. Because the aspect ratio is $\delta \approx 1$, the problem can be reduced to 2D and requires the solution of the perturbation density, $n_H'(x,z,t)$ by action of gravity waves on the mean background number density profile of haze particles, $n_H(z)$, which for simplicity is assumed to decrease exponentially with constant scale height H$_H$ with an assumed nominal value of 50 km. The linearized continuity equation for $n_H'(x,z,t)$ is:

(8) $$\frac{\partial n_H'(x,z,t)}{\partial t} + \frac{\partial [n_H(z)u'(x,z)]}{\partial x} + \frac{\partial [n_H(z)w'(x,z)]}{\partial z} = 0$$

The horizontal gravity wave velocity can be computed from the perturbation linear continuity equation for the nitrogen atmosphere mass density, $\rho(z)$,

(9) $$\frac{\partial u'(x,z)}{\partial x} + \frac{\partial w'(x,z)}{\partial z} - \frac{w'(x,z)}{H} = 0 \; ; \quad \text{where} \quad \frac{1}{\rho}\frac{d\rho}{dz} = -\frac{1}{H}$$

yielding for $u'(x,z)$,

(10) $$u'(x,z) = \frac{m}{k}w'(x,z) + \frac{w'(x,z)}{ikH}$$

with $w'(x,z)$ from Eq. (3). The time derivative in Eq. (8) is approximated by the period $2\pi/\Omega$ = 1 Pluto day, although in practice the growth of $n_H'(x,z,t)$ will be limited by the time it takes for its amplitude to become nonlinear, which certainly occurs when $n_H'(x,z,t) \sim n_H(z)$. Equation (9) divided by $n_H(z)$ reduces to

(11) $$\frac{n_H'(x,z,t)}{n_H(z)} = -\frac{2\pi}{\Omega n_H(z)}\left[\frac{\partial [n_H(z)u'(x,z)]}{\partial x} + \frac{\partial [n_H(z)w'(x,z)]}{\partial z}\right]$$

The solution is illustrated in Fig. 26 and shows ~ 25 distinct layers, with ~ 7 layers below 25 km. The layer spacing is a function of the vertical wavelength given to zeroth order by $\lambda_z \sim 2\pi u_0/N\delta$, where N is fixed by the background atmosphere so only $u_0$ and $\delta$ are individually unconstrained, except that $\delta$ <1.005 to have a propagating gravity wave for large $\lambda_y$ and in concert $u_0$ and $\delta$ are constrained observationally by the vertical spacing of haze layers.

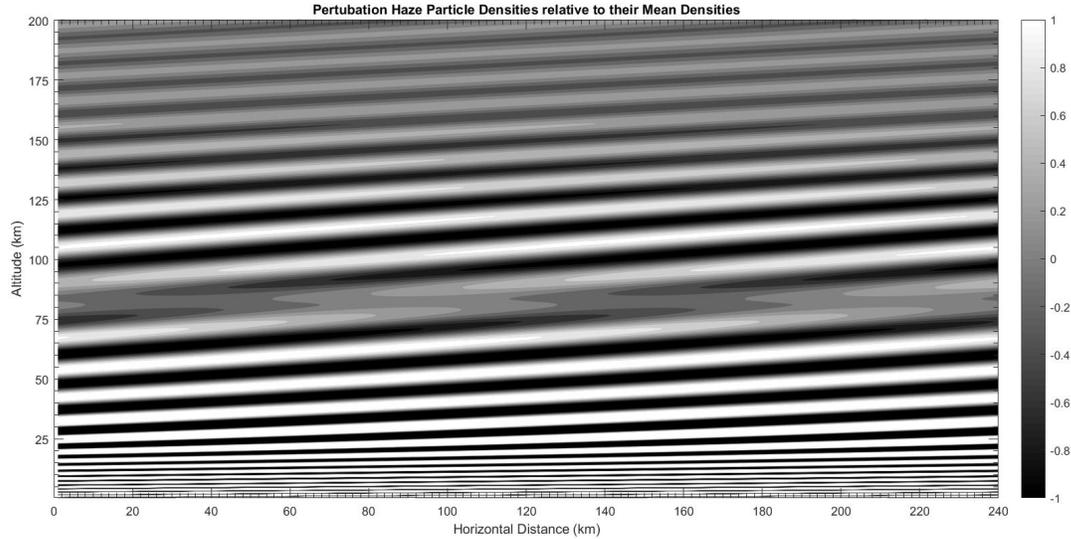

Figure 26 The perturbation haze particle densities, $n_H'(x,z,t)$, relative to the mean densities, $n_H(z)$, the left hand side and the solution of Eq. (11). Light areas correspond to large haze particle densities and dark areas to low densities. Shown are two horizontal wavelengths in the x direction, with the Fourier mountains at 0, 120, 240 km approximately and with maximum vertical forcing velocity at ~ 90 and 210 km. Background atmosphere is shown in Fig. 24.

Fig. 26 indicates that haze layering is not as distinct at ~ 75 km as at other altitudes. This is related to the fact that $N(z)$ has its minimum value in this region and thus $\lambda_z$ attains a maximum value in this region. In this region $n_H'(x,z,t)$ never exceeds the cutoff amplitude of $n_H(z)$. According to our theory for the input quantities in Fig. 24, the most distinct haze layers are located between 20-75 km and 100-125 km with the layering fading by the upper boundary at 200 km in Fig. 26. The calculation was carried out to 400 km but shown only up to 200 km to better display the layering. Above 250 km, one should include molecular viscosity and heat conduction that will eventually dissipate the gravity wave energy. The trend shown in Fig. 26 continues above 200 km with gradual fading out of the layering evident at lower altitudes, even without molecular viscous dissipation.

As shown in Fig. 26 for observers viewing in the +y direction the haze layers have a westward tilt (tilt to the left) consistent their upward propagation and energy transport. Note that observers viewing in the –y direction would see the layers tilted to the right. But if observers were viewing the model haze layers in the negative x direction, perpendicular to the y,z plane, they would see the layers as parallel to the surface. For other viewing directions, the degree and sign of the perceived tilt would depend on the viewing direction.

In spite of the theory's success in accounting for key features of the haze layers observed in Pluto's atmosphere, it should be regarded as semi-quantitative as no predictions have been made yet of the scattered light intensities associated with total haze particle densities $n_{HT}(x,z,t) = n_H(z) + n_H'(x,z,t)$.

The properties of gravity waves invoked to explain the haze layering may be compared with the those of the buoyancy waves inferred by Hubbard et al. (2009) in Pluto's middle atmosphere (150-300 km), from the observation of the 2007 March 18 stellar occultation of P445.3 by Pluto. They applied scintillation theory to the observed stellar light fluctuations at two wavelengths to infer gravity wave properties. They focused on two waves consistent with the gravity wave dispersion relation and the measured vertical wavelength of ~10 km: one with a horizontal wavelength of ~ 1000 km which is incompatible with our theory as orographic forced waves cannot propagate through the lower atmosphere where it is evanescent. The other wave had a horizontal wavelength of 60 km, similar to our preferred wavelength of ~ 120 km. Hubbard et al. (2009) concluded that the power spectrum as a function of vertical wavenumber for the latter wave was much more consistent with a saturated gravity-wave spectrum than a Kolmogorov spectrum. This is consistent with the gravity wave hypothesis in the present work that weak winds blowing over topography can reach saturation amplitudes close to the surface while propagating vertically and remaining saturated to altitudes as high as haze layering is observable.

The density fluctuations ($\rho' / \bar{\rho}$) retrieved by Hubbard et al. (2009) have amplitudes of ~|0.01| to be compared with the saturated wave amplitudes of ~|0.008| in the present model at the same altitudes (these are fluid density fluctuations, not shown, to be distinguished from the haze number density fluctuations of eq. 11). While Hubbard et al. (2009) inferred the atmosphere to remain sub-adiabatic in the presence of gravity waves, the present model invokes saturated gravity waves that render the atmosphere adiabatic where the wave's negative temperature gradient reaches a maximum. These local adiabatic regions are spaced about 10 km apart and thin, at most on the order of the Fresnel scale equal to ~3.2 km for 2007 occultation data and probably not observable. Wave saturation prevents the wave amplitude from growing with altitude. Consequently there is no need to invoke viscosity and thermal heat conduction to damp the wave below 250 km as Hubbard et al (2009) did. Wave saturation creates a diffusive turbulence with an effective diffusivity of $D \approx ku^4 / 2HN^3\delta^3$ (Lindzen, 1981, his Eq. 17). At 200 km for the present model wave parameters, this

yields D ~7 m² s⁻¹ which compares well with the Hubbard et al. (2009) viscous diffusivity of 6 m² s⁻¹. Furthermore, the time constant to transport energy over 1 vertical wavelength is one order of magnitude smaller than the viscous damping time constant up to 200 km.

## 8. Discussion and implications of atmospheric haze

Pluto's atmosphere is hazy from its surface up to altitudes of more than 200 km. Such an extended haze, at altitudes up to hundreds of km, is surprising at Pluto, because the atmospheric temperature is too high for condensation of hydrocarbons or nitriles at these altitudes except in the bottom ~15 km of the atmosphere. The detached haze layer in the upper atmosphere of Titan, near ~520 km altitude, similarly forms in a region where the temperature is too high for condensation of hydrocarbons or nitriles, and where the 1-10 μbar atmospheric pressure is similar to that where extended haze forms at Pluto. The processes that form the detached haze layer at Titan (Lavvas et al. 2009; Lavvas 2010) are proposed in Section 6 to form also the extended haze at Pluto. These processes involve low energy electron collisions with nitriles (HCN and $HC_3N$) initiating the formation of negative ions $CN^-$ and $C_3N^-$ and eventually leading to very large negatively charged macromolecules that attract positive ions (Vuitton et al.,2009), forming aerosols that grow, settle downwards and eventually coagulate collisionally into aggregates.

Sources of condensation nuclei in the Pluto atmosphere include particles mobilized by winds, atmospheric ions that create condensation nuclei, and infalling Kuiper Belt dust particles (Poppe 2015), which are presumed to be predominantly water ice. The infalling Kuiper Belt dust may yield a downward flux of water molecules, each of which acts as a nucleation site, as large as $\sim 1 \times 10^5$ cm⁻²s⁻¹. This may be the dominant source of condensation nuclei in the atmosphere.

Since formation of extended haze at Pluto is not easily understood unless there is an ionosphere, the New Horizons observations of extended haze in the atmosphere provide observational evidence supporting the existence of a Pluto ionosphere. The rate of ionization in Pluto's atmosphere is only about 12 times lower than that at Titan, and there are abundant sources of condensation nuclei, so that similar processes as those that produce Titan's detached haze layer may explain also the extended haze at Pluto.

The extended Pluto haze appears to be globally distributed and formed into ~20 thin (~few km), nearly horizontal layers (typically spaced by ~10 km) that extend horizontally over more than 1000 km in some cases. Extended haze is detected above the limb of Pluto, in both approach (low solar phase angle) images and departure (high solar phase angle) images, over both sunlit and night side limb, and at northern, equatorial and southern latitudes. The well-ordered, globally extensive layering is consistent with a quiet atmosphere at Pluto and low wind speeds. Such a quiet, close to globally uniform atmosphere at Pluto is what would be expected of an atmosphere whose pressure and composition are buffered by extensive surface ice-covered areas, principally $N_2$, with minor amounts of $CH_4$ and CO ices. Solar-induced sublimation of surface ices drives transport to colder, lower surface regions, buffered by subsequent condensation which constrains pressure variations in the atmosphere above the first ½- scale height to small values, $\Delta p/p$ < 0.2% for a surface pressure ~10 μbar (Young, 2012). With only small pressure variations in the atmosphere on a global scale, horizontal winds are expected to be weak, only on the order of meters/second. Nevertheless, such weak winds are sufficient to generate atmospheric gravity waves as they blow over Pluto's rugged terrain.

A model of orographic forcing of gravity waves in Pluto's atmosphere is presented in section 7 accounting for the thickness and spacing of extended haze layers and the extent of haze layers up to about 200 km altitude. In this model the vertical oscillations of an air parcel in a gravity wave lead to compressions and rarefactions, increasing and decreasing the density of aerosol particles, and thereby increasing and decreasing scattered light intensity. The observed contrast of a few to several percent in layer brightness would imply wave growth to non-linear amplitude in this model as shown in Fig. 26. The model predicts that layering becomes indistinct near an altitude ~75 km where the atmospheric Brunt Vaisala frequency reaches its minimum value. These features are in accord with observed layering of the Pluto extended haze. We note that the lowest atmospheric haze (up to 15 km altitude), which is also the brightest haze, forms in a region where hydrocarbons are super-saturated and the atmosphere is cold enough for direct condensation, unlike the case for extended haze. This lowest haze can form independently of any gravity waves. The maximum of haze brightness is in general reached several km above the limb of Pluto (as best illustrated in Figs. 4 and 5).

The haze in Pluto's atmosphere and the haze layering from visible imaging are not spatially uniform over the surface. The haze is brighter by a factor 2 to 3 over the northern latitude limb than over the equatorial limbs. The greater haze brightness over northern latitude limb indicates a greater line-of-sight opacity of haze particles in the north by about a factor of 2, as shown by comparing Fig 2 with the solar illumination model of Fig 19 which assumed a spatially uniform haze distribution. Although the haze is not as bright over equatorial latitudes, layering is more distinct over the equator, as illustrated in Figs. 1, 2, and 6; also see Fig. 11 comparing high resolution I/F profiles over northern and equatorial latitudes obtained over a short time span of four minutes.

A search for temporal variation of haze brightness was performed by comparing haze images over the same locations on Pluto and various times over time spans up to 5.43 hours. No motion of layers is detected in these comparisons, but changes in brightness I/F are seen versus height over the same geographic locations at different times. The brightness changes at the lowest altitudes correlate to solar phase angle, indicating a strongly forward scattering phase function for haze particles, but changes in the shape of I/F profiles (e.g., Figs 9b and 9d) indicate temporal variations in haze particle line-of-sight opacity.

Haze was detected in visible images over the limb of Pluto in both approach (low phase angle) and departure (high phase) images from the New Horizons flyby, spanning back scattering and forward scattering geometries respectively, but also spanning a 19 hour time interval over various locations on Pluto for the observation sequences given in Table 1. The scattered light from Pluto haze is about as bright as light scattered from Pluto surface at phase angle 148°. Haze is brighter than Pluto surface at higher phase angles, but the Pluto surface is much brighter than the haze at low phase angles.

Two phase functions for Pluto haze at visible wavelengths were compiled (Table 4) using only haze observations over the day side limb at similar northern latitudes (>40°). The phase function of Pluto haze at peak I/F is similar to that of cometary dust and that of Saturn's G-ring dust. A model of Mie scattering by 0.5 μm spheres (Fig. 20) reproduces this peak I/F phase function in the visible but gives UV extinction too low by an order of magnitude.

On the other hand, the visible wavelength phase function at higher altitude is notably different, lacking a backscatter lobe, although it has a similar forward scattering peak. The higher altitude phase function (Table 4) is similar to that found for Titan aerosols (Lavvas et al. 2010). A Titan-like model, of scattering by aggregate particles of bulk radius 0.15 μm and primary particle radius 0.01 μm (Fig. 21), reproduces the UV extinction and the visible forward scattering peak without a backscatter lobe. We note that the peak I/F phase function of Table 4 characterizes mainly haze particles in the lowest scale height, whereas the UV solar occultation senses mainly haze extinction at higher altitudes up to 300 km. We infer that the peak I/F phase function of Table 4 characterizes mainly particles at the bottom of the atmosphere where direct condensation may occur, forming relatively large and spherical particles. At higher altitude, the Pluto haze phase function is well-fitted by a Titan-like aggregate particle model.

Haze particles form and settle out of the atmosphere, onto Pluto's surface, rapidly over a time scale of days to weeks (Fig. 23). The rate at which haze particles settle out onto the surface is governed by production at high altitudes, which is ultimately limited by the methane photolysis rate (Gladstone et al. 2016), yielding an estimated haze particle deposition rate on the surface of $\sim 1 \times 10^{-14}$ g cm$^{-2}$s$^{-1}$. For an average haze particle radius of 0.5 μm and density of 1 g cm$^{-3}$, a haze particle layer of total geometric cross section area equal to the surface area would be deposited in ~200 years. Such a surface scattering layer would have ~unit optical depth, sufficient to alter optical properties. However, Pluto's atmospheric surface pressure varies seasonally (Hansen & Paige 1996; Young 2013; Hansen et al. 2015) and may drop sufficiently to halt haze production. Haze production peaks near 0.5 microbar; if the surface pressure drops below this, then haze deposition may stop before the accumulated particles can amount to even one monolayer. Moreover, at even smaller surface pressures, settled haze particles on the dark and non-icy surface regions become exposed to solar and interplanetary UV and solar wind, and these particles would undergo rapid photolysis, destruction and/or blackening before Pluto's atmosphere re-generates. In this picture, the dark regions on Pluto, like Cthulhu, can remain dark despite rapid settling of haze particles out of the atmosphere. On the other hand, haze that settles onto bright icy regions would be buried by freshly condensed volatiles and possibly protected from the UV and solar wind radiation processing. These haze particles may survive on Pluto's surface through its annual cycles and be slowly processed into tholins by cosmic rays. The alternative to this picture, that Pluto's atmosphere remains opaque to UV during the Pluto year, would imply that haze particles settle continually onto the surface and rapidly accumulate to an optically thick surface layer within thousands of years. These particles would not be processed into tholins except by cosmic rays, and the striking albedo contrasts on Pluto, with very bright and dark regions, would be difficult to understand.

**Acknowledgements.** This work was supported by NASA under the New Horizons Project. We thank W. B. Hubbard for a helpful review of the manuscript.